# Spatially inhomogeneous delithiation in LiNiO$_2$ positive electrode: the effect of X-rays dose

F. La Porta[1], L. Barthe[1], A. Beauvois[1], G. Wittmann[2], V. Briois[1,3], A. Iadecola[4,5], S. Belin[1]

[1] Synchrotron SOLEIL, L'Orme des Merisiers, Départementale 128, 91190 Saint-Aubin, France
[2] Technische Universität München, Arcisstraße 21, 80333 München, Germany
[3] Centre National de la Recherche Scientifique, CNRS UR1, France
[4] Réseau sur le Stockage Electrochimique de l'Energie (RS2E), CNRS FR3459, 80039 Amiens, France
[5] Sorbonne Université, CNRS, Physicochimie des Electrolytes et Nanosystèmes Interfaciaux, CNRS UMR8234, Paris, France

*corresponding authors: francesco.la-porta@synchrotron-soleil.fr; stephanie.belin@synchrotron-soleil.fr; Antonella.iadecola@synchrotron-soleil.fr

## Abstract

Operando synchrotron X-ray techniques have become essential tools for investigating rechargeable batteries as they provide real-time insights into electrochemical processes. However, the high brilliance of synchrotron radiation can alter the electrochemical mechanisms within the battery, thereby compromising the reliability and reproducibility of operando measurements. In this study, we introduce a novel methodology that directly correlates the local X-ray dose with the Ni$^{4+}$/Ni$^{3+}$ redox activity in a LiNiO$_2$ positive electrode. Full-field transmission X-ray absorption spectroscopy imaging (FFI-XAS) is employed to probe the charge-compensation mechanism during lithium extraction at the micrometre scale, using two beam configurations with different focal distances (far-focus and near-focus). While the spatially averaged XAS spectra exhibit sluggish reaction, regions exposed to lower dose rates exhibit the expected electrochemical evolution. This contrast enables the identification of a dose threshold for reliable operando measurements. This approach establishes a practical dose limit and provides spatially resolved insight into beam-induced effects, offering both a diagnostic framework and a pathway toward more reliable operando experiments.

## 1. Introduction

Rechargeable batteries are identified as a crucial asset for the transition from fossil fuels to renewable energy sources [1]. Understanding the complex phenomena occurring in the bulk electrodes or at the interfaces in *operando* conditions, *i.e.* during the battery working, is necessary to improve their electrochemical properties, such as capacity retention and lifespan, and to develop safer devices [2]. Synchrotron-based *operando* techniques have been widely applied to investigate the structural, chemical and morphological changes during cycling in different battery configurations, ranging from model devices to commercial batteries [3]. Beyond well-established techniques for battery studies, such as X-ray diffraction (XRD) [4], [5], [6] X-ray diffraction computed tomography (XRD-CT)[7], [8], [9] and X-ray absorption spectroscopy (XAS)[7], [8], [10], [11], [12], [13], advanced techniques have also been developed with complex sample environments. Indeed, trapping of soluble polysulfides within the carbon was observed at the nanoscale using *operando* scanning X-ray microscopy (STXM) [9] in Li-S batteries. Moreover, near ambient pressure X-ray photoelectron spectroscopy (NAP-XPS) was essential to monitor the formation of the solid-electrolyte interface (SEI) in model Li-ion batteries [14], [15], [16]. Recently, grazing incident X-ray scattering (GIXRD) [17], [18] was demonstrated to be a powerful technique to detect the formation of the SEI circumventing the experimental restrictions required for NAP-XPS experiments.

While the previous studies yielded important insights, attention should be paid to potential interactions between X-rays and electrochemical processes, which could induce parasitic effects and result in misleading data interpretation [19]. One of the parasitic phenomena commonly induced by beam interaction relies on the formation of secondary electrons, exciting inner-shell electrons responsible for the Auger decay [20]. The resulting photoelectrons and Auger electrons initiate ionisation cascades, contributing to the radiolysis of organic components [20], [21] such as the polymeric binder [22], [23], [24], [25], [26], the cathode–electrolyte interphase CEI [20], [21], [27] and the electrolyte [28], [29]. The effects of such detrimental interactions can create inhomogeneities in the electrochemical reactions [30] or in the severe case inhibit it completely in the irradiated spot[19].

Although synchrotron *operando* experiments are becoming routine, only a few works were focused on the impact of the X-ray beam on the reaction mechanisms. Black and coworkers have investigated how various experimental conditions and analytical techniques influence $LiNi_{0.33}Mn_{0.33}Co_{0.33}O_2$ (NMC) and $LiFePO_4$ (LFP) cathode materials [20]. This work correlated the observed beam-induced effects with radiation-dose parameters, including incident energy, photon flux, and exposure time [20]. Jousseaume *et al.* have shown that beam damage is caused by the intermixed actions of dose and dose rate in NMC-graphite batteries, causing only mild modifications of the crystalline structure at low dose, but inducing artificial phase transitions at high dose [19]. Through comparison of *ex situ*, *operando*, and *in situ* data, Blondeau *et al.* have demonstrated that beam exposure delays the electrochemical reactivity, thus biasing the correct interpretation of the conversion mechanism in the illuminated spot [29]. All these studies revealed that the threshold dose strongly depends on the chemistry of active material of the electrodes and the electrolyte formulation. Consequently, although determining the dose threshold looks like mandatory to ensure reliability of the results, this is still extremely challenging. This process often relies either on preliminary experiments in which measurements with attenuated incident beam are compared with high dose measurements, or on the use of raster-scanned measurements across the electrode. These approaches help mitigate or avoid undesirable effects at the irradiated spot while maintaining a reasonable signal-to-noise ratio in the collected data [19].

Defining experimental conditions or protocols to mitigate or avoid undesirable phenomena in battery studies is therefore of primary importance, especially with the advent of $4^{th}$-generation synchrotrons with higher brilliant X-ray sources. In this paper, we introduce a novel experimental method based on full-field transmission Quick-XAS imaging (FFI-XAS) enabling the estimation of multiple X-ray dose effects within a single *operando* XAS experiment. Our approach takes advantage of the time resolution and the chemical sensitivity - provided by FFI-XAS - combined with micrometric spatial resolution [31], [32], resulting in hyperspectral experiment. To validate our method, a benchmark electrode made of $LiNiO_2$ active material was cycled in half-cell configuration, as its electrochemical behaviour as well as its charge compensation mechanism have already been studied in our group [8], [33], [34].

Moreover, an analytical method was proposed to retrieve the relationship between the local dose experienced at each pixel and the corresponding XAS spectrum. This enables the investigation of the reactions in a wide range of doses during a single experiment [20], [21] and allows us to determine the dose threshold at which the sample starts to be affected by the beam irradiation. In addition, this approach accounts for the intrinsic spatial inhomogeneity of the incident beam, providing an accurate tool that can be readily adapted to any type of X-ray experiment.

Overall, the proposed approach showcases the effects of a broad spectrum of doses while preserving spatial information on beam-induced alterations in the battery, thereby offering practical guidelines for X-ray synchrotron studies aiming to minimize or avoid beam effects in batteries.

# 2. Experimental

## Materials

Al-supported $LiNiO_2$ electrode was provided by BASF; it consisted of 10 µm layer of $LiNiO_2$ active material mixed with a binder (PVDF) and conducting carbon ($LiNiO_2$:PVDF:Carbon in the weight ratio 94:3:3) which was deposited and calendared on 20 µm Al foil. Its nominal area capacity was 1.0 mAh/cm². For the *operando* experiments, disks of 14 mm of diameter were cut from the foil. The separator was 18 mm diameter Celgard 2500, and 14 mm diameter lithium foils (MTI-Corp) were adopted as the negative electrode in half-cell configuration. The electrolyte was provided by E-lyte, we have used 50 µl of LP57 (E-lyte) (1M $LiPF_6$ in EC:EMC 3:7 v/v). All the cell components were dried at 80°C in a Buchi vacuum oven overnight before use.

## Electrochemistry

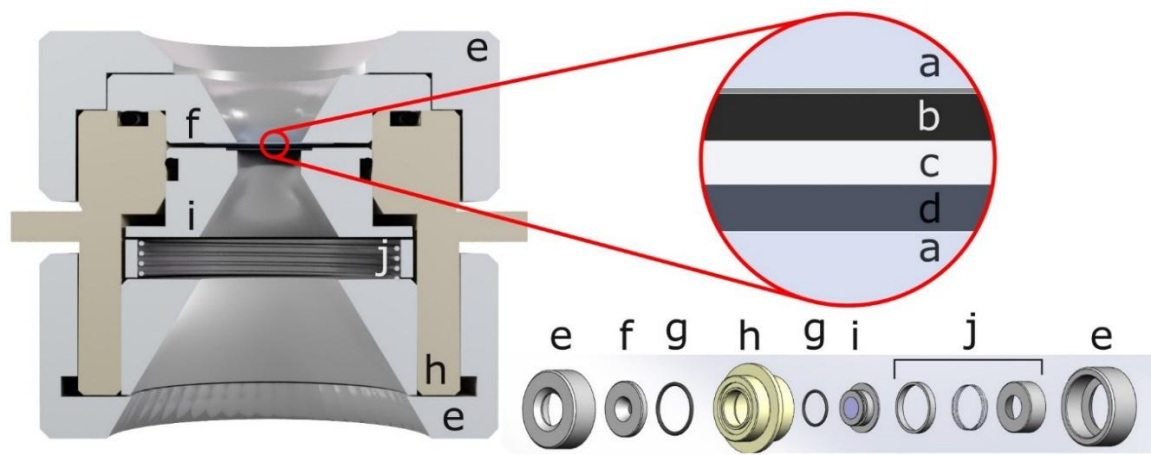

*Figure 1: Schematic view of the ROCK cell. a) Diamond windows, b) $LiNiO_2$ on aluminium foil, c) Celgard separator, d) Lithium foil, e) external flanges, f) Molybdenum frame on the positive electrode side, g) o-rings, h) PPS body, i) Molybdenum frame on the negative electrode side, j) spring system*

The *operando* XAS experiments were carried out using a custom-built cell, hereafter referred to as ROCK cell. The general concept for this cell is adopted from a classic modified coin cell configuration, and it is shown in Figure 1. The cell consists of two X-ray transparent windows mounted onto the main body using two threaded lids (Figure 1e). The main body is made of hydrophobic PolyPhenylene Sulfide PPS (Figure 1h). The sizes of the windows (Figure 1a) are smaller than the electrode stack (Figure 1b,c,d) in order to provide electrical contact on both electrodes across the metallic part (Figure 1f,i). The mechanical pressure is applied to the battery stack using a multilayer wave spring (Borelli), providing a nominal pressure of 4.5 bar onto the stack (Figure 1j). The air tightness of the cell is ensured by the O-ring gaskets (Figure 1g). A 6 mm diameter and 145 µm-thick polycrystalline CVD diamond plates were used as X-ray transparent windows and brazed on metallic molybdenum (Figure 1a). The CVD diamond plates were hard enough to withstand homogeneous pressure of up to 8 bar before breaking and are air-tight due to the brazing manufacturing. Diamond was preferred as a nonconductive window to prevent parasitic electrochemical reactions while enabling X-rays transmission. Cell assembling is done by mounting the piston and spring on the "front" side of the cell first. Then the battery is assembled inside the now recessed surface of the piston, which facilitates concentric alignment of the electrodes. Finally, the cell is sealed by mounting the plate window element on the "back" side of the cell.

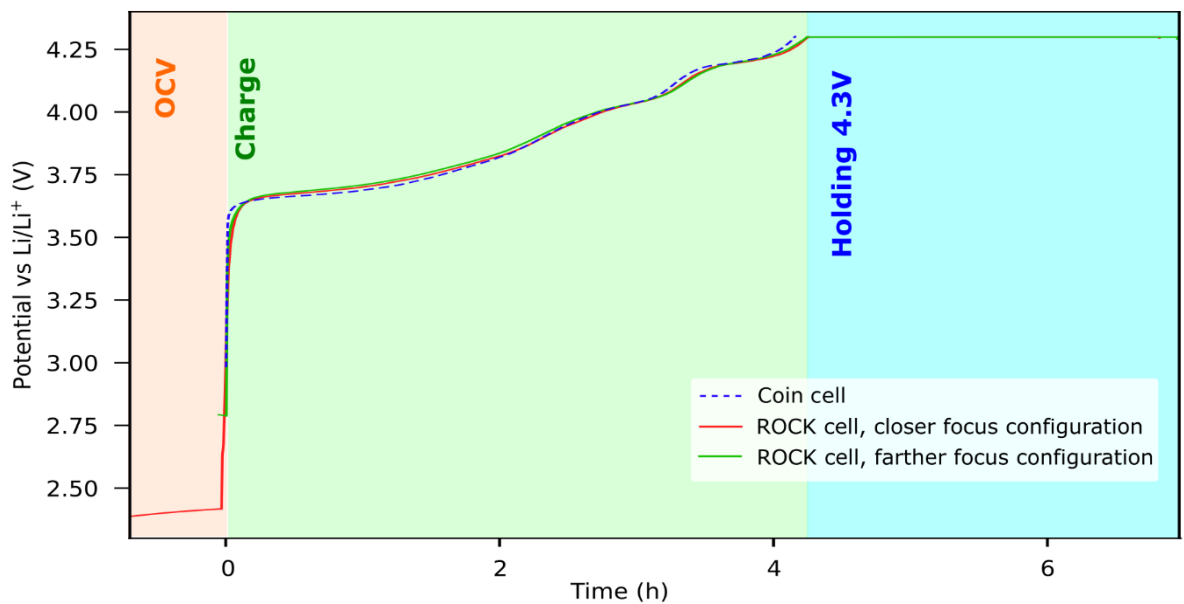

*Figure 2: First delithiation of LiNiO$_2$ against Li metal performed at C/4 (1 Li$^+$ deintercalated in 4 hours). The galvanostatic charge in the coin cell (dashed blue line) is compared to those obtained in the ROCK cell (red and green lines).*

Two ROCK cells have been tested and their electrochemical performances compared to those of 2032 coin cell (Figure 2, dashed blue line). In all cases, the galvanostatic charge was performed by applying a constant current of 0.375 mA/h corresponding to the C-rate of C/4 (1 Li$^+$ deintercalated in 4 hours) and the cutoff voltage was set to 4.3 V vs Li$^+$/Li to avoid further electrolyte degradation. The galvanostatic curve during charge stems from the phase changes of the LiNiO$_2$, as already reported and discussed in the literature and the Li$^+$ extraction is balanced by the oxidation of Ni from 3+ to 4+ [8], [33], [34],.

After the voltage of 4.3 V was achieved, the potential was held at 4.3 V for approximately 3 hours, in order to record additional FFI-XAS data at steady state conditions.

## X-ray absorption spectroscopy

X-ray absorption spectroscopy experiments, both FFI-XAS and non-spatially resolved Quick-XAS, were performed at the ROCK beamline of synchrotron SOLEIL (France) [31], [35]. The storage ring was operated in top-up mode at 500 mA during the whole experiment. The X-rays were emitted by a 2.81 T SuperBend source [35] and horizontally focused by a toroidal mirror to the experimental hutch at 32.68 m from the source. The horizontal beam size is controlled by translating the sample along the focusing axis inside the experimental hutch from 32.68 m to 26 m upstream of the source. The experiment was conducted by placing the cell at two different focal positions of the toroidal mirror, thus resulting in two different horizontal beam sizes. In addition, the sagittal bent M2B mirror located after the ROCK monochromator was used to vertically focus the beam. A moderate vertical focusing, as the one typically used for conventional Quick-XAS monitoring of electrode materials [32] has been used, spreading the beam over a vertical area around 1.0 ± 0.2 mm. In summary, the cell was positioned at 28 meters from the X-rays source, being the beam ~ 3.28(H) x 1.21(V) mm$^2$, referred hereafter as 'farther focus configuration' and at 29.9 m, with beam size ~ 1.75(H) x 0.92(V) mm$^2$ referred as 'closer focus configuration'.

The FFI-XAS was acquired using the set-up configuration described by Briois *et al*. [31], [32]. The X-rays photons were converted into visible light via a 100 μm-thick YAG:Ce scintillator (Crytur) placed at 2 mm from the ROCK cell and collected on an ORCA Flash4.0 V3 digital CMOS camera (Hamamatsu) through a 4X lens (Navitar 4X HR Plan Apo Objective). The camera's sensor array of 4.0 megapixels consisted of 2048(H)x2048(V) pixels, each of dimensions 6.5x6.5 μm$^2$, however, to improve the speed of the acquisition, for the imaging measurements, the dimensions were reduced to 2048(H)x748(V). Considering the magnification of the 4X lens, the sample-to-pixel size was 1.625 μm. A Si(111) Quick-XAS channel-cut monochromator with a Bragg angle of 13.467° and an oscillation amplitude of 1.51° was used to acquire[36], [37] the spectra at Ni K-edge (8333 eV) spanning the 8266–8878 eV energy range. In FFI-XAS experiment, the oscillation frequency of the monochromator was set to 0.09 Hz and 580 images of 2048×748 pixels were recorded and stacked in edf cube during the increasing angles of the monochromator oscillation, thus resulting in a three-dimensional data file (HC, hyperspectral cube).

In non-spatially resolved Quick-XAS measurements, the frequency of the monochromator was set at 2 Hz. Moreover, vertical and horizontal exit slits were used to define a beam size of 0.5 x 0.5 mm² to limit the vertical energy dispersion of the monochromator [32], [38]. The incident beam intensity ($I_0$), the transmitted intensity from the sample ($I_1$) and from the Ni reference foil ($I_2$) were measured simultaneously with three OKEN ionisation chambers filled with $N_2$.

## Experimental protocol and image post-processing

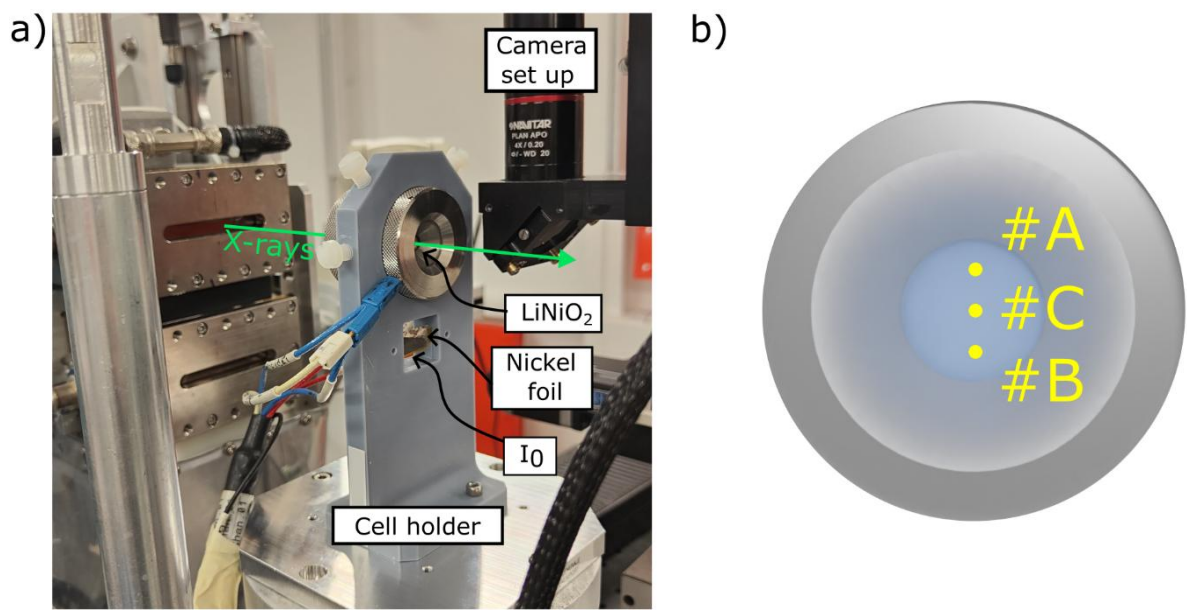

*Figure 3: a) Picture of the experimental setup. The ROCK cell is mounted on a sample holder (the direction of the beam is indicated by the green arrow). The Ni foil was placed on the same focal point of the LiNiO$_2$ electrode, and $I_0$ was measured by vertically translating the sample stage. The camera setup (on the right side) could be translated perpendicularly in and out of the beam direction. b) Graphical representation of the front view of the ROCK cell (piece f in Figure 1) with the spots used for collecting the FFI-XAS data (in yellow).*

The ROCK cell was placed on a custom-made holder, as shown in Figure 3a. The holder had an empty space below the cell enabling the collection of the flat-field ($I_0$) cubes on the CMOS camera and hosts

Ni foil, which is used for collecting a reference hyperspectral cube in the same sequence of acquisition of flat-field and sample cubes on the same area of the CMOS sensor. The recording of those reference cubes is mandatory for the post-data analysis to correct the spatial energy dispersion at each pixel of the image, as detailed in [32]. The cell holder was fixed on a translational stage to align the different targets (LiNiO$_2$ electrode, Ni foil, I$_0$). The camera was mounted on a second translational stage to align the beam with the center of the CMOS sensor and to switch from FFI-XAS to Quick-XAS configuration [32].

At the beginning of each experiment, 20 Quick-XAS hyperspectral cubes (QXAS-HC) were collected on the incident beam I$_0$ for flat field correction and on the Ni foil. Moreover, 120 QXAS-HC and a non-spatially resolved Quick-XAS spectra (QXAS-IC) were collected at OCV in order to measure the initial oxidation state of the LiNiO$_2$ electrode. Then, the LiNiO$_2$ electrode was delithiated up to 4.3 V, concomitantly to the acquisition of QXAS-HC, resulting in 1400 QXAS-HC over 4 hours (1 QXAS-HC every 11 seconds). During the charge, the X-rays were continuously illuminating the sample at the point #C (Figure 3b).

Once the charge was achieved, the voltage's cell was held at 4.3 V for 3 hours. In the meanwhile, 20 QXAS-HC were collected for both I$_0$ and the Ni foil reference, for a total acquisition time of 3.67 minutes, respectively. Then, the cell was again moved into the beam, and additional 120 QXAS-HC and QXAS-IC spectra were recorded at three different positions: the continuously irradiated position #C, and two fresh points, namely #A and #B (Figure 3b). Each acquisition lasted 22 minutes and 10 minutes, for QXAS-IC and 120 QXAS-HC, respectively.

The post-processing of the hyperspectral cubes was done using the Jupyter notebooks developed at the beamline [31], [32]. The first step implies averaging all the QXAS-HC obtained in steady-state conditions (for example at OCV) into a single QXAS-HC to enhance the signal-to-noise ratio (S/N). The 20 raw data cubes of the I$_0$ and Ni foil and the 120 cubes of the OCV and during the voltage hold were therefore averaged into 4 different cubes (1 for I$_0$, 1 for Ni foil, 1 for the OCV and 1 for the voltage holding). The Lambert-Beer law was then applied to obtain the corresponding absorption spectra in each pixel of the image for the electrode material and the Ni foil. The pixels of the images were then binned 4x4 to improve the S/N leading to a final binned pixel size of 6.5 x 6.5 µm$^2$. In this condition, the images of the 'farther focus' configuration results in 512x187 pixels with a corresponding field of view (FoV) of 3.28x1.21 mm², while a region of interest of 400x187 pixels was defined for the 'closer focus' configuration (FoV 2.60x1.21 mm²). Each image was associated to an energy selected by the monochromator of the beamline. The pre-edge was subtracted using a linear function, while the post-edge was modelled with a polynomial function of degree 2. The energy calibration was done using the first inflection point of the Ni foil reference spectra to 8333 eV at each binned pixel of the image. The resulting energy shift depends on the pixel position and defines the energy shift map, which is applied to each pixel absorption spectra to realign correctly all the QXAS-HC.

Principal Component Analysis (PCA) and Multivariate Curve Resolution with Alternative-Least Squares (MCR-ALS) were applied to describe the phase evolutions of the LiNiO$_2$ upon delithiation. The PCA algorithm is used to determine the number of principal components (Cps) capturing the variance of the dataset. Then, the spectra of pure species coexisting in the phase mixture and their relative concentration are reconstructed from the LiNiO$_2$ dataset using the MCR-ALS algorithm. This analysis was performed using MCR-ALS GUI 2.0 developed by Tauler *et al.* on the MATLAB platform [39]. The fit was done imposing the non-negativity constraints on the spectra and on the concentrations and forcing the total concentration of all the components to be always equal to 1 (closure constraint). All QXAS-HC can be analysed using MCR-ALS, which enables the determination of the concentrations of each species on every binned pixel, thereby creating a concentration map.

## Dose estimation analysis

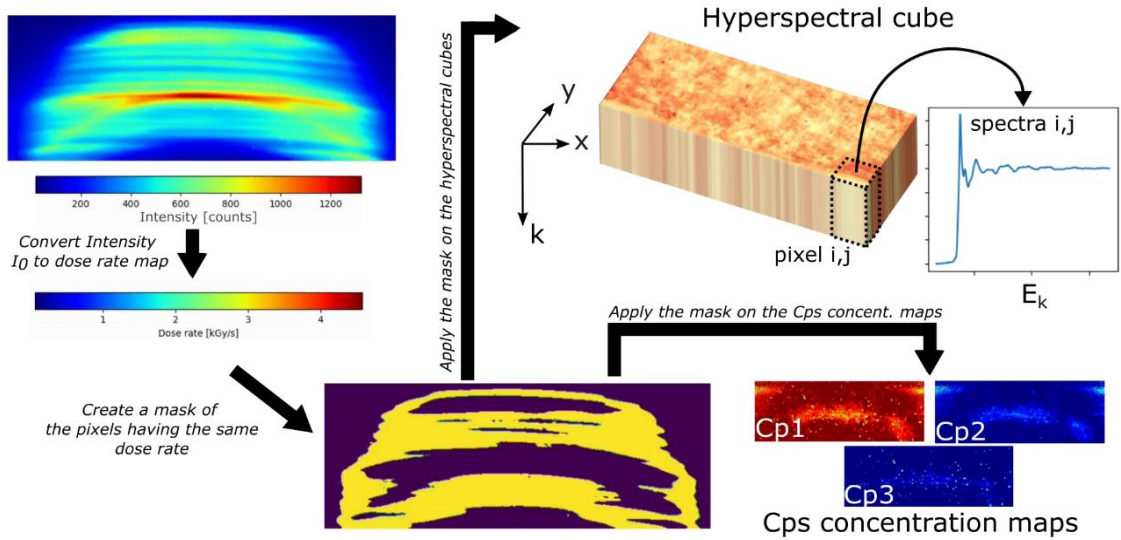

*Figure 4: Description of the procedure for the dose estimation, the definition of the dose rate maps and its application to the Cps concentration maps. The flat-field HC ($I_0$) was converted first into a flux map following the Eq.1, and then into a dose rate map (Eq.2). The dose rate map is then used to create a mask of dose rate, that can be applied on the QXAS-HC or directly on the Cp concentration maps.*

The main steps of the analytical process to calculate the dose rate map were depicted in Figure 4. The total flux after the first ionization chamber ($I_0$) was measured using a calibrated diode resulting in $1.041 \cdot 10^{12}\ ph.s^{-1}$ at 8390 eV. Before LiNiO$_2$ however, a significant photons fraction is absorbed by all the upstream components of the cell: one CVD diamond window, the lithium counter electrode, the Celgard separator, and the electrolyte. The total absorption of the upstream elements is calculated using the Jupyter script provided by Jousseaume et al. [19] (the details of the calculations are shown in the SI), and it accounts for 21% attenuation of the incoming total flux. The total photon flux $F_t$ on the electrode is then $8.22 \cdot 10^{11}\ ph.s^{-1}$. Being the intensity of the X-ray beam spatially inhomogeneous, the dose is expected to be also inhomogeneous. Therefore, it is important to define a dose map to consider the local effect of the dose on the homogeneous LiNiO$_2$ electrode material. For this purpose, the intensity recorded by each pixel of the $I_0$ cube (in counts) was converted into photon flux ($ph \cdot s^{-1}$), applying the following relation:

$$F_{i,j,k}[ph \cdot s^{-1}] = \frac{I_{i,j,k}}{I_{tot,k}} F_t[ph \cdot s^{-1}] \qquad \text{Eq. 1}$$

where $F_{i,j,k}$ is the flux of the $(i,j)$ pixel of the $k$ image, $I_{tot,k}$ is the sum of the intensities of all the pixels of the image $k$ and $F_t$ is the total photon flux. Each image $k$ corresponds to an energy selected by the monochromator during the measurement.

Once the flux map was obtained, it can be converted into the dose rate map using the following formula [19], [20], [40]:

$$DR_{i,j,k}\ [Gy\ s^{-1}] = \frac{E_k[J] \cdot (1 - T_k)}{A_{pix}[m^2]\ \tau[m]\ \rho\left[\frac{kg}{m^3}\right]} F_{i,j,k}[ph \cdot s^{-1}] \qquad \text{Eq. 2}$$

where $DR_{i,j,k}$ is the dose rate for each $(i,j)$ pixel of the image $k$, $E_k$ is the photon energy corresponding to the image $k$, $T_k$ is the transmitted intensity, $A_{pix}$ is the pixel area, $\tau$ is the absorber thickness and $\rho$ is the absorber density. The unit of $DR_{i,j,k}$ is $[Gy\ s^{-1}]$ that is equivalent to $[m^2 s^{-3}]$. It is important to recall that this approach can be applied because the commercial Al-supported LiNiO$_2$ cathode is very homogeneous in the chemical composition and its thickness does not vary significantly within the FoV of the camera.

The $DR_{i,j,k}$ is a 3D tensor, the energy-averaged dose rate map $AV(DR_{i,j})$ related to each QXAS-HC can be calculated:

$$AV(DR_{i,j})[Gy\ s^{-1}] = \frac{1}{k}\sum_{1}^{k}DR_{i,j,k}[Gy\ s^{-1}]$$

Eq. 3

Once the energy-averaged dose rate map $AV(DR_{i,j})$ is calculated, we can create a mask considering specific dose ranges. The dose rate mask can be either directly applied to the QXAS-HC, or to the Cp concentration maps to correlate the dose rate with the Ni oxidation state. Moreover, the dose rate mask can easily convert into a dose mask by multiplying it by the duration of the beam exposure.

# 3. Results

## Farther focus configuration

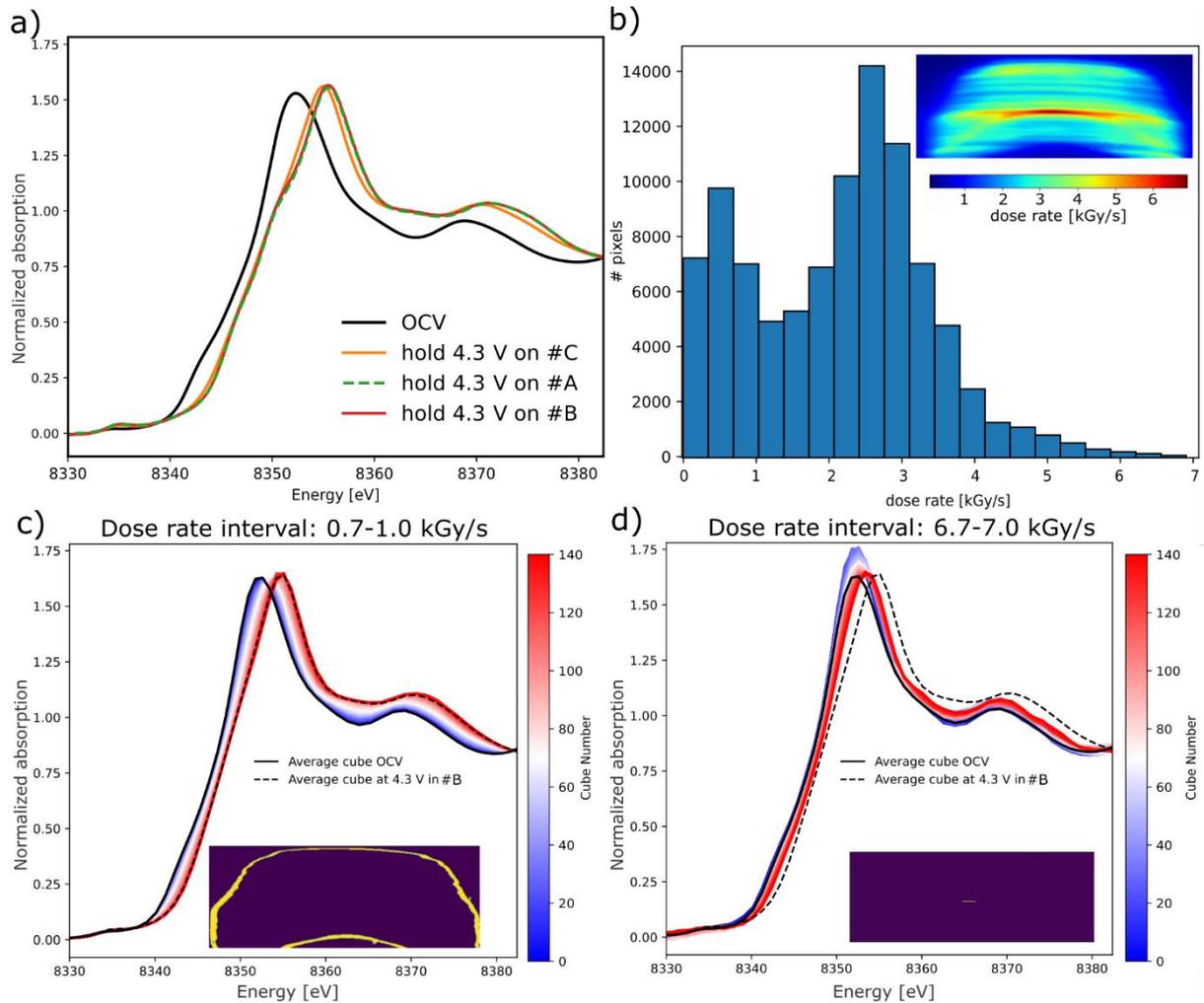

*Figure 5: Operando XAS experiment in farther focus configuration. a) QXAS-IC spectra collected at OCV and at 4.3 V at the end of the holding at the continuously irradiated position #C, and in two fresh points #A and #B. b) Histogram of number of pixels as a function of the dose (step of 0.3 kGy/s), the dose rate map is shown in the inset. c) Evolution of the averaged QXAS-HC corresponding to 0.7 - 1 kGy/s dose rate during the charge, the corresponding dose rate mask is shown in the inset (yellow). d) Evolution of averaged the QXAS-HC corresponding to 6.7 - 7 kGy/s dose rate during the charge, the corresponding dose rate mask is shown in the inset (yellow).*

Upon delithiation, LiNiO$_2$ undergoes several structure phase transitions, with a subsequent collapse of the c-axis of the pristine rhombohedral structure and gliding of the NiO$_2$ slabs [41]. Its structural evolution occurs concomitantly to the redox activity of the Ni sites, where Ni is oxidized from the initial 3+ to the final 4+ state to accommodate the charge neutrality. The change of the Ni oxidation can be easily detected by XAS spectroscopy, which reveals a shift of the Ni K-edge toward higher energy, as described already by Jacquet *et al.* [8]. In particular, the authors have applied PCA and MCR-ALS algorithms to the *operando* XAS spectra, and 3 distinct components were reconstructed corresponding to 3 different Ni electronic configurations and local environments. The initial component corresponds to LiNiO$_2$ pristine phase, with a formal oxidation state of Ni$^{3+}$, and characterized by a distorted NiO$_6$ octahedral configuration with four 1.92 Å Ni–O bonds and two 2.04 Å Ni–O bonds [34], [41]. The second intermediate component reached a maximum concentration of approximately 50% at mid-charge, and it exhibited a reduced distortion in the NiO$_6$ octahedra, indicating partial delithiation [8], [42]. The third component, where the NiO$_6$ octahedra become symmetric and the final formal oxidation Ni$^{4+}$ is achieved. The results of this study are particularly interesting because the beam effects were carefully addressed by probing multiple positions on the sample and attenuating the incident beam at any single spot. As a result, the reported measurements were only marginally affected by beam damage and can be reliably used as a reference for comparison with the results presented in this work.

The first *operando* experiment was conducted in the farther focus configuration, by placing the cell at 28 m from the ROCK's source, and the beam size was 3.28(H) x 1.21(V) mm$^2$ (FWHM). Figure 5a shows the non-spatially resolved XANES spectra collected at open circuit voltage (OCV) and at 4.3 V (during holding) at three different sample positions. The edge position of the XANES spectra recorded at 4.3 V is shifted to higher energy compared to the one at OCV, as expected with the activation of the Ni$^{4+}$/Ni$^{3+}$ redox couple during the delithiation. By carefully comparing the three spectra at 4.3 V, the one irradiated all along the charge (#C) has a different shape around 8335 eV compared to those in the fresh points #A and #B. This difference would result from the higher integrated radiation dose received at position #C compared to the other positions.

As discussed previously, the incoming beam is not homogeneous, and therefore the dose rate is spatially inhomogeneous [23], [37]. Therefore, the local dose rate for each binned pixel of the image was calculated using the approach described in the previous section. The number of pixels as a function of the dose rate intervals is reported in the Figure 5b, while the dose rate map is shown in the inset. The pixel distribution was bimodal with a first peak at lower dose rate (corresponding to the edges of the beam) and another one at a higher dose rate (corresponding to the central part of the beam).

Two dose rate intervals were selected, one between 0.7-1.0 kGy/s (low dose rate) and the other one between 6.7-7.0 kGy/s (high dose rate), and the corresponding QXAS-HC recorded over the pixels receiving the selected dose rate were averaged. Figure 5c-d show the averaged Ni K-edge XANES spectra upon charge corresponding to the two dose rate ranges. For the lower dose rate (0.7-1.0 kGy/s), the last XANES spectrum recorded at 4.3 V in point #C overlaps with the one collected in the fresh point #B, suggesting that the delithiation of the LiNiO$_2$ was fully achieved in both cases. In contrast, the evolution of the XANES spectra corresponding to the higher dose rate (6.7-7.0 kGy/s), clearly shows that the reaction was not fully completed in the considered pixels.

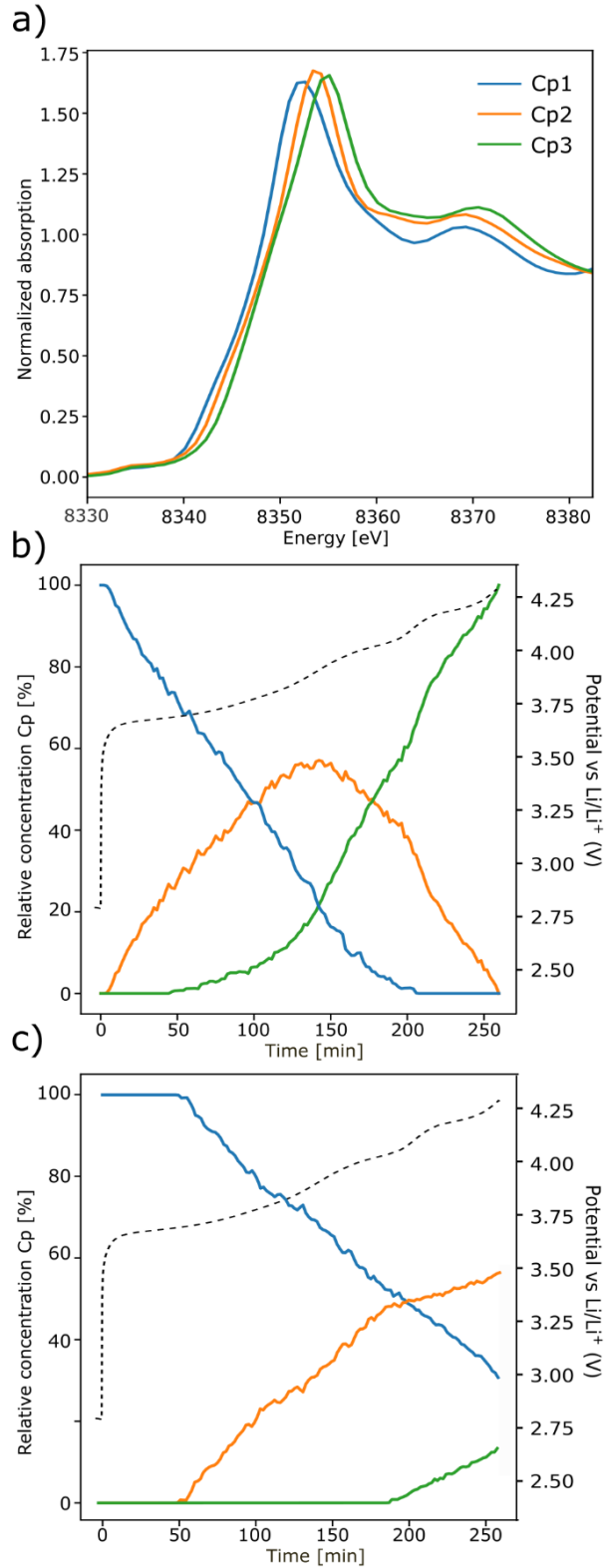

*Figure 6. a) MCR-ALS reconstructed spectra for the dataset collected during delithiation at position #C using the mask 0.7-1.0 kGy/s. The Cp1 is depicted in blue, the Cp2 in orange and the Cp3 in green. All the QXAS-HC spectra were reconstructed as linear combination of these Cps. b) Evolution of relative concentration of the three Cps considering the dose rate between b) 0.7-1.0 kGy/s and c) 6.7-7.0 kGy/s, along with the galvanostatic charge (dashed line).*

In order to get further insights on the effect of the dose rate on the Li$^+$ extraction, we have applied MCR-ALS analysis to the averaged QXAS-HC (Figure 5c) measured for the dose rate of 0.7 - 1.0 kGy/s.

The details of the chemometric analysis are reported in the Figures S1-S4. The reconstructed spectra of the pure species involved during the charge and their relative concentrations are shown in Figure 6b. The first reconstructed spectra Cp1 corresponds to the spectra of $LiNiO_2$ at OCV (as shown in the S5a); the second principal component Cp2 corresponds to an intermediate phase of the material, while the third principal component Cp3 correspond to the fully delithiated $Li_0NiO_2$ and is comparable to the spectra recorded at 4.3 V in the fresh #B position (in Figure S5b).

The relative concentration of Cp1 measured by FFI-XAS (Figure 6b) decreased linearly throughout the charging process, becoming negligible above 4.1 V. On the other hand, the relative concentration of Cp3 started to increase around 3.7 V, and it reaches 100% at the end of the charge. The intermediate component (Cp2) exhibited a concentration peak at 58% at 3.8 V and then it gradually diminished. In agreement with previous results [8], we can conclude that a dose rate between 0.7 and 1.0 kGy/s is sufficient to avoid the beam effect on the redox reaction mechanism.

The QXAS-HC experiencing the higher dose between 6.7-7.0 kGy/s (Figure 5d) were averaged and then deconvolved using the previously reconstructed MCR-ALS spectra (Figure 6a) as input basis. In this case, the concentrations of Cp1 and Cp2 are modified only after 50 minutes, above 3.7 V, and these components persist along the charge. On the other hand, the component Cp3 appears above 4.1 V and coexists with the other two components until the end of the charge. While the applied current would force the $Li^+$ extraction from the $Li_{1-x}NiO_2$ electrode during the charge, an opposite force would slow down the Ni redox reaction.

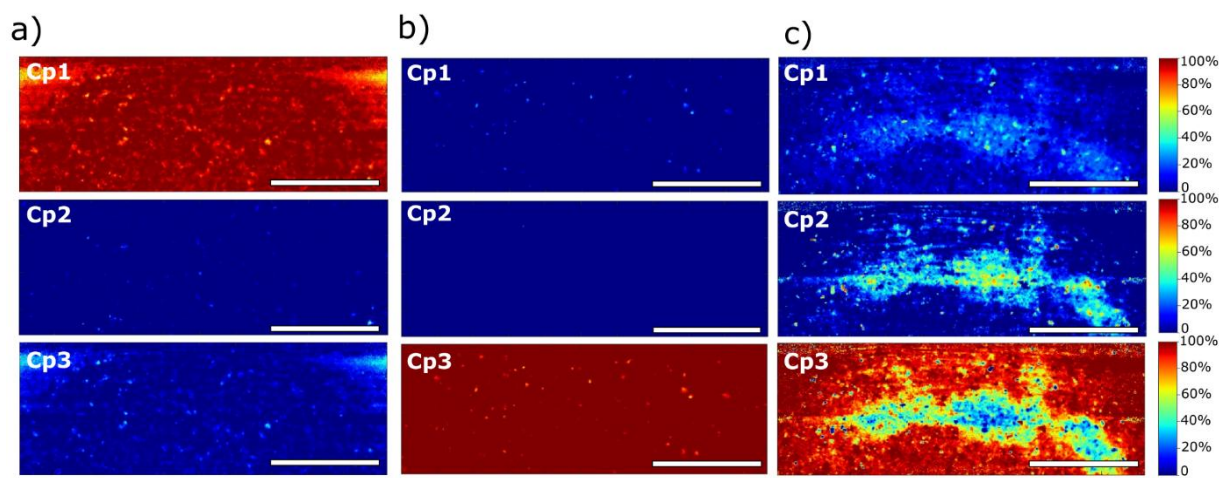

*Figure 7: Concentration maps of the three reconstructed components Cps in farther focus configuration. At OCV (a), at 4.3 V in a fresh position #B (b) and at 4.3 V in continuously irradiated spot #C (c), the latter with higher x-rays dose accumulated during the charge process. The scale bar is 1 mm.*

To verify this hypothesis, we have decomposed the QXAS-HC collected at OCV, at 4.3 V in the continuously irradiated spot #C and at 4.3 V in the fresh spot #B, using the 3 reconstructed components (Cps presented in Figure 6a) to retrieve their relative map concentration (Figure 7). For QXAS-HC, three maps were calculated, and they represent the spatial concentration distribution of Cps. The R-factor maps of the MCR-ALS fit are shown in Figure S6.

At OCV (Figure 7a), the dominant species is Cp1, corresponding to the pristine state. Because the number of incident photons is very low in the two small regions at the top of the image, the resulting fits are noisier and less accurate. At 4.3 V, the relative concentration of Cp3 is 100% in the fresh spot #B (Figure 7b), whereas the reaction is incomplete in the spot #C. In particular, a clear beam effect is visible at the center of these concentration maps (Figure 7c): the delithiation is not fully achieved in this region, as Cp1 (10–20%) and Cp2 (30–50%) are still detected. Cp3 is predominantly found at the edges of the beam where low dose rate was deposited on the electrode. We can thus conclude that in the continuously irradiated #C position, the reaction of the electrode is inhomogeneous because of the high dose rate related to X-ray beam.

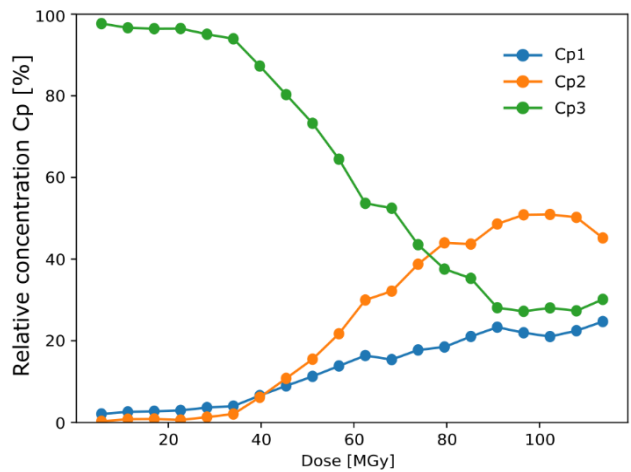

Figure 8: Relative concentration of the reconstructed components Cps obtained in spot #C at 4.3V in farther focus configuration as a function of the total exposure time. The dose corresponds to the dose rate multiplied by the cumulative exposure time.

Finally, our objective was to determine the dose threshold at which the beam begins to alter the delithiation reaction. For this purpose, we have calculated the relative concentration of the reconstructed Cps for different local dose in spot #C at 4.3 V and the results are shown in Figure 8. The concentration values for each dose were obtained in 2 steps: first the dose map was created multiplying the dose rate by the exposure time (local dose), then a mask of the selected dose was applied to each concentration map (Figure 7c) and then averaged. Ideally, at the end of the first charge, only Cp3 species should be present. This assumption is valid for integrated dose up to 35 MGy; above such value, the relative concentration of the reconstructed Cp2 and Cp1 increases at the expense of Cp3. This method, therefore, establishes 35 MGy as the critical dose above which the beam substantially impacts the electrochemical process for the studied $LiNiO_2$ electrode materials.

## Closer focus configuration

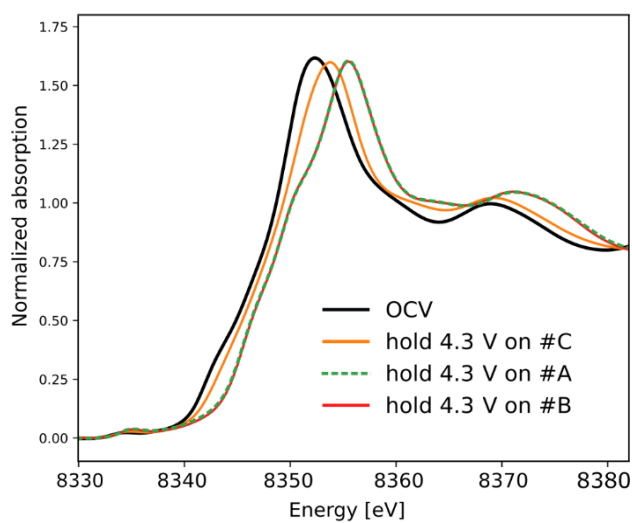

Figure 9: QXAS-IC spectra in the closer focus configuration at OCV, at 4.3 V at the end of the holding at the continuously irradiated position #C, and in two fresh points #A and #B.

To deepen our understanding of the beam effect, we have repeated the experiment by placing the ROCK cell at a longer distance from the X-ray source (29.9 m). In this configuration, the photon flux remains unchanged while the beam size is reduced, resulting in a higher photon density per unit area on the LiNiO$_2$ electrode. The non-spatially resolved QXAS-IC spectra acquired at OCV and at the voltage held at 4.3 V in the three positions are shown in Figure 9. In this configuration, the dose is expected to be significantly higher than in the previous case, thus worsening the inhibition/delay of the electrochemical reaction in the irradiated region of the electrode. Indeed, in the "closer focus" configuration, the edge energy of the spectrum recorded in the continuously irradiated point #C is red-shifted compared to those in the fresh points #A and #B, while only small differences were observed in the "farther focus" configuration (Figure 5a). This indicates that the closer focus configuration irradiation condition has a more severe effect on the oxidation of the Ni cations than the farther focus one, despite the galvanostatic curve is not influenced by the higher dose (Figure 2a). This apparent discrepancy can be explained because only 2.6% of the total cathode area is exposed to the X-ray, thus its influence on the macroscopic galvanostatic measurement is negligible.

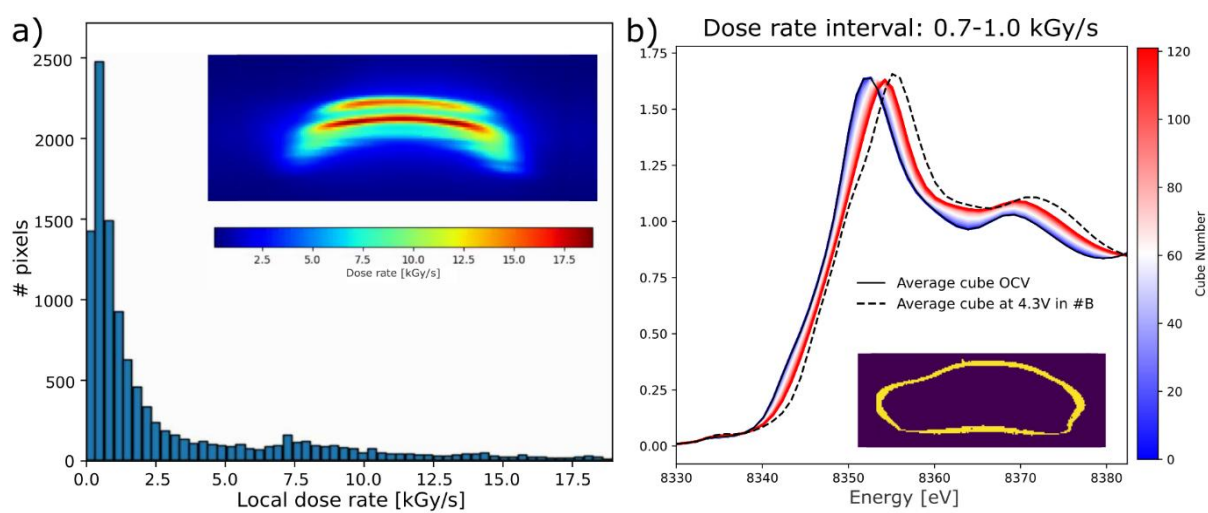

*Figure 10: Operando XAS experiment in closer focus configuration. a) Histogram of number of pixels as a function of the dose (step of 0.3 kGy/s), the dose rate map is shown in the inset. b) Evolution of averaged QXAS-HC corresponding to 0.7 – 1.0 kGy/s dose rate during the charge, the corresponding dose rate mask is shown in the inset (yellow).*

The dose rate distribution and the dose rate map in the FoV of the camera are shown in the Figure 10a. The maximum dose rate of 18.5 kGy/s is higher than in the farther focus configuration (7 kGy/s), the beam is concentrated in a smaller area. Unlike previous configuration (Fig 5b), the pixel distribution is essentially unimodal with most of the pixels receiving a lower dose rate than in the farther focus configuration, and they are located on the edges of the map, while higher doses are concentrated in the center.

To compare the results between the closer and farther focus configurations, the same dose rate between 0.7-1.0 kGy/s was considered for the pixels in the spot #C. All QXAS-HC corresponding to this dose rate were binned, and the resulting XANES spectra in the #C point during the delithiation along with the spectrum collected in the fresh point #B are shown in the Figure 10b. While considering the same dose rate, the last *operando* XANES spectrum is still shifted at lower energy with respect to the spectrum acquired at 4.3 V in #B. This means that, even at the lowest dose rate, the Ni oxidation was not fully achieved in the continuously irradiated #C point. This finding demonstrates that the dose rate cannot be simply considered as the unique criterion to avoid the beam radiation effect in battery studies.

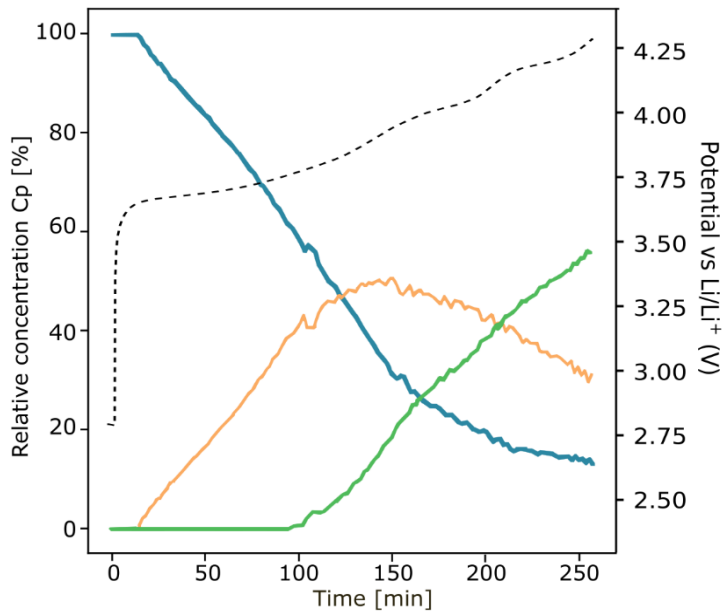

*Figure 11: Evolution of relative concentration of the three Cps considering the dose rate between b) 0.7-1.0 kGy/s in closer focus configuration along the galvanostatic curve (dashed line). For sake of comparison, we have used the same Cps as in the farther focused configuration at 0.7-1.0 kGy/s (Figure 6a).*

To monitor the reaction homogeneity upon delithiation for dose rate between 0.7 and 1.0 kGy/s, the QXAS-HC spectra (Figure 10b) were deconvolved as a linear combination of the previously reconstructed Cps (Figure 6a). Considering this input basis, the evolution of their relative concentration is shown in Figure 11 (details of the calculation in Figure S7-S8). The concentration of Cp1 decreased to 13% at the end of the delithiation, the Cp2 reached its maximum (51%) at 3.9 V and it is still present at 4.3 V. The Cp3 appeared above 3.76 V (almost at half-charge) and increased until it reached 56%. The spectrum at the end of charge, therefore, did not correspond to a fully delithiated $Li_0NiO_2$, but rather an intermediate $Li_{0.6}Ni_{0.4}O_2$ composition, as all three reconstructed Cps are still present at the high voltage.

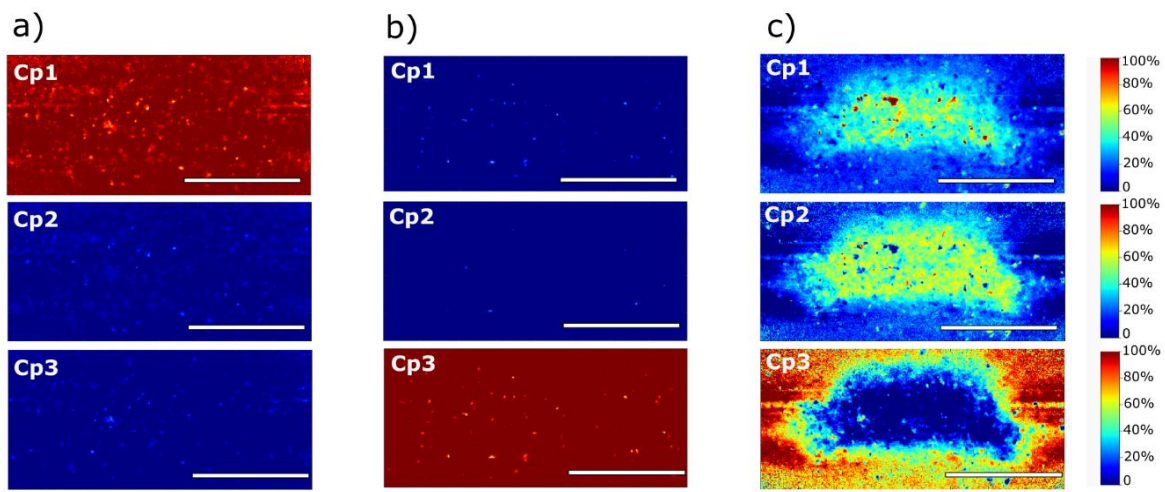

*Figure 12: Concentration maps of the three reconstructed components Cps in the case of closer focus beam configuration at OCV (a), at 4.3 V in #B (b), and at 4.3 V in #C (c). The white scale bar is 1 mm.*

The reconstructed Cp spectra shown in Figure 6a) were used to deconvolute the QXAS-HC data obtained at OCV and at the end of holding at 4.3 V. The concentration maps are presented in Figure 12 (the R-factor maps in Figure S9). As expected, the concentration map at OCV shows only the presence of Cp1 (Figure 12a). On the other hand, only Cp3 is detected in the fresh spot #B (Figure 12b) at the end of the holding at 4.3 V, as observed in the farther focus beam configuration. However, the three reconstructed species are present in the concentration maps acquired in the spot #C at 4.3 V (Figure 12c). In particular, Cp1 is predominant at the centre of the image (where the dose is higher), and Cp3 is mainly observed at the edge of the beam, where the number of photons is significantly lower. We can conclude that in the closer focus configuration, the delithiation reaction is deeply inhibited.

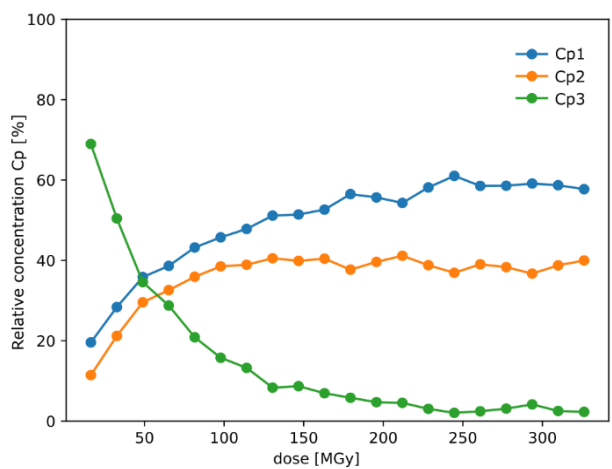

*Figure 13: Relative concentration of the reconstructed components Cps in closer focus configuration obtained in spot #C at 4.3V as a function of cumulated dose. The dose corresponds to the dose rate multiplied by the cumulative exposure time.*

We have calculated the dose accumulated in the point #C during the Li extraction and the voltage hold in the closer focus configuration, in order to identified also in this case a threshold value. The maximum local dose is 320 MGy, more than 2.5 times higher than the maximum local dose calculated in the farther focus beam configuration (120 MGy). Figure 13 shows the evolution of the Cps relative concentrations as a function of the cumulated dose received in the spot #C. It is interesting to note that even at the lowest dose of 16 MGy, the contribution of Cp3 remains below 80% and it rapidly decreases as dose increases, unlike the previous case (Figure 13). In addition, the concentration of Cp1 is always higher than that of Cp2, which was not the case for the farther focus configuration.

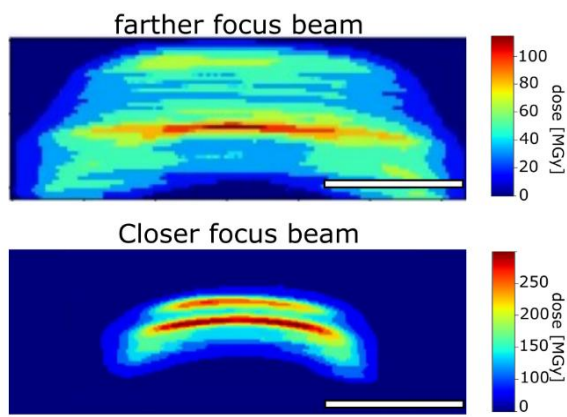

*Figure 14 : Dose map for the farther (top) and closer focus (bottom) configuration. The white scale bar is 1 mm.*

The observed differences at high doses can be explained by considering that the X-ray beam also influences adjacent regions [43]. Figure 14 shows the local dose maps for both beam configurations: in the case of a closer focus, the areas with high doses and the ones with low doses are a few µm far, and the beam effects can spread quickly from the higher dose region to a lower dose one. In our present study, each pixel represents a small portion of the sample (6x6 µm²), and we assume that the beam effect on a single pixel does not influence its neighbouring pixels. However, this assumption fails when considering the physicochemical nature of the investigated system, because, particularly at high doses, the damage can spread, as will be discussed in more details afterwards.

## 4. Discussion

We have investigated the homogeneity of the Ni redox activity during the electrochemical delithiation of $LiNiO_2$ oxide, used as positive electrode. Rather than gradually attenuating the intensity of the incident X-ray beam, we have performed two distinct experiments by varying the dose rate applied to the battery, i.e. by moving *the in situ* cell along the focusing direction of the toroidal optic shaping the horizontal beam size at the ROCK beamline. In both configurations, we have observed an incomplete redox reaction at the end of the charge in the continuously irradiated region of the $LiNiO_2$ electrode, considering the spectra recovered by merging all the HC pixels (Figure S10). However, a different mechanism is observed when we consider the regions of the electrode experiencing low and high dose rate, corresponding to 0.7–1.0 kGy/s and 6.7–7.0 kGy/s, respectively.

In the farther focus beam configuration, the beam footprint is significantly larger (3.3 × 1.2 mm²), and approximately 7,000 pixels were exposed to the low dose rate (0.7–1.0 kGy/s). The evolution of the corresponding XANES spectra follows the expected reaction pathway, from $Ni^{3+}$ to $Ni^{4+}$, once the cutoff voltage of 4.3 V is reached. At high dose rate, the electrochemical reaction becomes sluggish, leading to an incomplete redox conversion within the continuously irradiated region of the $LiNiO_2$ electrode. Nevertheless, the $Ni^{4+}$ final state is still observed in freshly irradiated spots.

A different behaviour is observed in the closer focus beam configuration, with smaller beam size (1.8 × 0.9 mm²). In this case, approximately 1,500 pixels were irradiated even at the low dose rate, and the reconstructed Ni redox activity is strongly delayed. This observation indicates that the dose rate alone cannot account for the beam-induced parasitic effects, and that additional parameters—such as beam size, irradiated area, or cumulative dose—must be considered. As reported in the literature, the electrochemical reactions are sensitive to the radiation dose: a higher dose can sluggish or even inhibit the reaction [3], [19]. These apparent difference in behaviour between the two configurations stems on the fact that beam-induced parasitic effects are not confined to the irradiated region. Some chemical reactions can be locally generated under the beam, but their effects can spread to the entire electrode. Two notable examples are electrode volume expansion and diffusion of damaged electrolyte molecules. The degradation of the binder may also induce swelling in the vicinity of the irradiated region, damaging a larger volume [23]. Interestingly, it has been proven for many polymers that the radiolysis of the binder may have a specific dose onset below which the polymer is not affected [26]. The second effect is the radiolysis of the molecules composing the electrolyte, in particular on organic electrolytes [21]. The damaged molecules generated in an area irradiated with high dose can spread to the surrounding regions, affecting the charge exchange mechanism of the nearby electrode active particles [29]. Moreover, X-ray irradiation can also induce defects on inorganic electrodes, thus influencing the reaction mechanisms [21]. For these reasons, the kinetics of the reaction of each pixel cannot be considered only dependent on the radiation dose calculated with the method described:

the reaction occurring at each pixel may also be influenced by the neighbouring regions, especially if they received a high dose.

For the farther focused configuration, the number of pixels receiving low dose is almost 4 times higher than in the closer focused configuration, where also the higher dose pixels are in proximity to the lower dose ones. The diffusion effects generated on heavily irradiated pixels can therefore easily influence the reaction of low dose pixels. In the closer focused configuration, the diffusion of the detrimental reactions is predominant for all the pixels, and therefore, the kinetics of the redox activity cannot be uniquely associated with the radiation dose, but proximity with higher dose pixels must be considered. Overall, our study demonstrates that several parameters must be considered when designing a successful *operando* battery experiment. The calculation of dose alone is not sufficient to guarantee the absence of parasitic beam-induced effects: the incident beam homogeneity and size must also be taken into account.

# 5. Conclusions

In summary, we have proposed a novel *operando* method to measure the X-rays beam effects on the electrochemical reactions of battery materials. This approach exploits the combined chemical and spatial resolution of FFI-XAS, allowing comprehensive characterization of the incident beam properties. By using a non-uniform tunable beam size, we can address the homogeneity and reliability of delithiation in $LiNiO_2$ positive electrode across a broad range of doses in a single experiment, and the large number of spectra collected from each pixel allows for more robust statistical analysis. The homogeneity of delithiation associated to the Ni redox reaction was therefore studied in two experimental configurations as a function of the local dose on each pixel of the detector camera. *Operando* XAS data were compared with *in situ* spectra collected on fresh spots to quantitatively assess on the homogeneity of the reaction and its correlation with the dose.

The key point of our method lies on the construction of a local mask to selectively extract spectra within a specific dose range, which correspond to the beam-free electrochemical reactions. By carefully curating the spectral dataset and subsequently applying the MCR-ALS chemometric approach, we have reconstructed the $Ni^{4+}/Ni^{3+}$ redox activity while excluding regions significantly affected by the parasitic beam-driven reactions.

Our *operando* FFI-XAS approach, including the definition of dose masks, provides a powerful framework to capture and rationalize the consequences of different experimental configurations in battery studies, enabling the design of more robust and reliable *operando* experiments.

## Acknowledgements

This work was supported by European Union's Horizon 2020 programme through grant agreements BIG-MAP (No 957189) and by a public grant overseen by the French National Research Agency (ANR) as part of the "Investissements d'Avenir" program (reference: ANR-10-EQPX-45). Furthermore, the authors want to thank the SOLEIL Synchrotron staff, in particular Claude Menneglier and Marie Andrae of the Detector Group for the insights about the scintillator and the camera; Emmanuel Farhi, Olga Roudenko of the GRADES group and Andrew King of PSICHE beamline for the interesting discussions about the analysis of the hyperspectral cubes. The authors would also like to thank Stéphanie Blanchandin of the chemistry laboratory of SOLEIL for her help.

# 6. Bibliography

[1]   A. Kalair, N. Abas, M. S. Saleem, A. R. Kalair, and N. Khan, "Role of energy storage systems in energy transition from fossil fuels to renewables," *Energy Storage*, vol. 3, no. 1, Feb. 2021, doi: 10.1002/est2.135.

[2]   J. Huang, S. T. Boles, and J. M. Tarascon, "Sensing as the key to battery lifetime and sustainability," Mar. 01, 2022, *Nature Research*. doi: 10.1038/s41893-022-00859-y.

[3]   A. P. Black *et al.*, "Synchrotron radiation based operando characterization of battery materials," Dec. 12, 2022, *Royal Society of Chemistry*. doi: 10.1039/d2sc04397a.

[4]   J. Sottmann, V. Pralong, N. Barrier, and C. Martin, "An electrochemical cell for operando bench-top X-ray diffraction," *J. Appl. Crystallogr.*, vol. 52, no. 2, pp. 485–490, 2019.

[5]   A. V Llewellyn, A. Matruglio, D. J. L. Brett, R. Jervis, and P. R. Shearing, "Using in-situ laboratory and synchrotron-based x-ray diffraction for lithium-ion batteries characterization: A review on recent developments," *Condens. Matter*, vol. 5, no. 4, p. 75, 2020.

[6]   K. Choudhary *et al.*, "Operando X-ray diffraction in transmission geometry at home from tape casted electrodes to all-solid-state battery," *J. Power Sources*, vol. 553, p. 232270, 2023.

[7]   G. Aquilanti *et al.*, "Operando characterization of batteries using x-ray absorption spectroscopy: advances at the beamline XAFS at synchrotron Elettra," *J. Phys. D Appl. Phys.*, vol. 50, no. 7, p. 74001, 2017.

[8]   Q. Jacquet *et al.*, "A Fundamental Correlative Spectroscopic Study on Li1-xNiO2 and NaNiO2," *Adv. Energy Mater.*, Nov. 2024, doi: 10.1002/aenm.202401413.

[9]   J. Nelson *et al.*, "In operando X-ray diffraction and transmission X-ray microscopy of lithium sulfur batteries," *J. Am. Chem. Soc.*, vol. 134, no. 14, pp. 6337–6343, 2012.

[10]  F. Yang *et al.*, "In Situ/Operando (Soft) X-ray Spectroscopy Study of Beyond Lithium-ion Batteries," *Energy & Environmental Materials*, vol. 4, no. 2, pp. 139–157, 2021.

[11]  B. Li *et al.*, "Decoupling the roles of Ni and Co in anionic redox activity of Li-rich NMC cathodes," *Nat. Mater.*, vol. 22, no. 11, pp. 1370–1379, 2023.

[12]  Y. Shirazi Moghadam *et al.*, "Unravelling the Chemical and Structural Evolution of Mn and Ti in Disordered Rocksalt Oxyfluoride Cathode Materials Using Operando X-ray Absorption Spectroscopy," *Chemistry of Materials*, vol. 35, no. 21, pp. 8922–8935, 2023.

[13]  S. Park *et al.*, "Irreversible Electrochemical Reaction at High Voltage Induced by Distortion of Mn and V Structural Environments in Na4MnV (PO4) 3," *Chemistry of Materials*, vol. 35, no. 8, pp. 3181–3195, 2023.

[14]  Y.-C. Lu *et al.*, "In situ ambient pressure X-ray photoelectron spectroscopy studies of lithium-oxygen redox reactions," *Sci. Rep.*, vol. 2, no. 1, p. 715, 2012.

[15]  J. Maibach *et al.*, "Probing a battery electrolyte drop with ambient pressure photoelectron spectroscopy," *Nat. Commun.*, vol. 10, no. 1, p. 3080, 2019.

[16]  I. Kallquist *et al.*, "Potentials in Li-ion batteries probed by operando ambient pressure photoelectron spectroscopy," *ACS Appl. Mater. Interfaces*, vol. 14, no. 5, pp. 6465–6475, 2022.

[17]  S. M. Bhaway *et al.*, "Operando grazing incidence small-angle X-ray scattering/X-ray diffraction of model ordered mesoporous lithium-ion battery anodes," *ACS Nano*, vol. 11, no. 2, pp. 1443–1454, 2017.

[18]  M. Bogar, I. Khalakhan, A. Gambitta, Y. Yakovlev, and H. Amenitsch, "In situ electrochemical grazing incidence small angle X-ray scattering: From the design of an electrochemical cell to an exemplary study of fuel cell catalyst degradation," *J. Power Sources*, vol. 477, p. 229030, 2020.

[19]  T. Jousseaume, J.-F. Colin, M. Chandesris, S. Lyonnard, and S. Tardif, "How beam damage can skew synchrotron operando studies of batteries," *ACS Energy Lett.*, vol. 8.8, pp. 3323–3329, 2023.

[20]  A. P. Black *et al.*, "Beam Effects in Synchrotron Radiation Operando Characterization of Battery Materials: X-Ray Diffraction and Absorption Study of LiNi0.33Mn0.33Co0.33O2 and LiFePO4 Electrodes," *Chemistry of Materials*, vol. 36, no. 11, pp. 5596–5610, Jun. 2024, doi: 10.1021/acs.chemmater.4c00597.

[21]  G. Leita and B. Bozzini, "Impact of space radiation on lithium-ion batteries: A review from a radiation electrochemistry perspective," Oct. 15, 2024, *Elsevier Ltd*. doi: 10.1016/j.est.2024.113406.

[22]  T. Coffey, S. G. Urquhart, and H. Ade, "Characterization of the effects of soft X-ray irradiation on polymers," 2002. [Online]. Available: www.elsevier.com/locate/elspec

[23]  C. Lim, H. Kang, V. De Andrade, F. De Carlo, and L. Zhu, "Hard X-ray-induced damage on carbon-binder matrix for in situ synchrotron transmission X-ray microscopy tomography of Li-ion batteries," *J. Synchrotron Radiat.*, vol. 24, no. 3, pp. 695–698, May 2017, doi: 10.1107/S1600577517003046.

[24]  L. A. Pesin, V. M. Morilova, D. A. Zherebtsov, and S. E. Evsyukov, "Kinetics of PVDF film degradation under electron bombardment," *Polym. Degrad. Stab.*, vol. 98, no. 2, pp. 666–670, Feb. 2013, doi: 10.1016/j.polymdegradstab.2012.11.007.

[25]  A. L. Sidelnikova *et al.*, "Kinetics of radiation-induced degradation of CF2- and CF-groups in poly(vinylidene fluoride): Model refinement," *Polym. Degrad. Stab.*, vol. 110, pp. 308–311, 2014, doi: 10.1016/j.polymdegradstab.2014.09.009.

[26]  S. A. Vaselabadi, D. Shakarisaz, P. Ruchhoeft, J. Strzalka, and G. E. Stein, "Radiation damage in polymer films from grazing-incidence X-ray scattering measurements," *J. Polym. Sci. B Polym. Phys.*, vol. 54, no. 11, pp. 1074–1086, Jun. 2016, doi: 10.1002/polb.24006.

[27]  R. Fantin, A. Van Roekeghem, J. P. Rueff, and A. Benayad, "Surface analysis insight note: Accounting for X-ray beam damage effects in positive electrode-electrolyte interphase investigations," *Surface and Interface Analysis*, vol. 56, no. 6, pp. 353–358, Jun. 2024, doi: 10.1002/sia.7294.

[28]  M. Schellenberger *et al.*, "Accessing the Solid Electrolyte Interphase on Silicon Anodes for Lithium-ion Batteries In-situ through Transmission Soft X-ray Absorption Spectroscopy," *Mater. Today Adv.*, vol. 14, p. 100215, Jun. 2022.

[29]  L. Blondeau, S. Surblé, E. Foy, H. Khodja, S. Belin, and M. Gauthier, "Are Operando Measurements of Rechargeable Batteries Always Reliable? An Example of Beam Effect with a Mg Battery," *Anal. Chem.*, vol. 94, no. 27, pp. 9683–9689, Jul. 2022, doi: 10.1021/acs.analchem.2c01056.

[30]  C. K. Christensen *et al.*, "Beam damage in operando X-ray diffraction studies of Li-ion batteries," *J. Synchrotron Radiat.*, vol. 30, no. Pt 3, pp. 561–570, Mar. 2023, doi: 10.1107/S160057752300142X.

[31]  V. Briois *et al.*, "Multimodal Insights of Regeneration Dynamics of Spent Bimetallic Catalysts by Full Field Hyperspectral Quick-EXAFS Imaging and Environmental Transmission Electron Microscopy," *ChemCatChem*, Sep. 2024, doi: 10.1002/cctc.202400352.

[32]  V. Briois *et al.*, "Hyperspectral full-field quick-EXAFS imaging at the ROCK beamline for monitoring micrometre-sized heterogeneity of functional materials under process conditions," *J. Synchrotron Radiat.*, vol. 31, no. Pt 5, pp. 1084–1104, Sep. 2024, doi: 10.1107/S1600577524006581.

[33]  I. Konuma *et al.*, "Unified understanding and mitigation of detrimental phase transition in cobalt-free LiNiO2," *Energy Storage Mater.*, vol. 66, p. 103200, 2024.

[34]  M. Bianchini, M. Roca-Ayats, P. Hartmann, T. Brezesinski, and J. Janek, "There and back again—the journey of LiNiO2 as a cathode active material," *Angewandte Chemie International Edition*, vol. 58, no. 31, pp. 10434–10458, 2019.

[35]  P. Brunelle *et al.*, "Development of a custom-made 2.8 T permanent-magnet dipole photon source for the ROCK beamline at SOLEIL," *J. Synchrotron Radiat.*, vol. 30, pp. 695–707, May 2023, doi: 10.1107/S1600577523002990.

[36]  V. Briois *et al.*, "ROCK: The new Quick-EXAFS beamline at SOLEIL," in *Journal of Physics: Conference Series*, Institute of Physics Publishing, 2016. doi: 10.1088/1742-6596/712/1/012149.

[37]  V. Briois *et al.*, "ROCK: the new Quick-EXAFS beamline at SOLEIL," in *Journal of Physics: Conference Series*, 2016, p. 12149.

[38]  M. Katayama, K. Sumiwaka, K. Hayashi, K. Ozutsumi, T. Ohta, and Y. Inada, "Development of a two-dimensional imaging system of X-ray absorption fine structure," *J. Synchrotron Radiat.*, vol. 19, no. 5, pp. 717–721, Sep. 2012, doi: 10.1107/S0909049512028282.

[39]  J. Jaumot, A. de Juan, and R. Tauler, "MCR-ALS GUI 2.0: New features and applications," *Chemometrics and Intelligent Laboratory Systems*, vol. 140, pp. 1–12, 2015.

[40]  A. Gimat, S. Schöder, M. Thoury, M. Missori, S. Paris-Lacombe, and A. L. Dupont, "Short- A nd Long-Term Effects of X-ray Synchrotron Radiation on Cotton Paper," *Biomacromolecules*, vol. 21, no. 7, pp. 2795–2807, Jul. 2020, doi: 10.1021/acs.biomac.0c00512.

[41]  L. De Biasi *et al.*, "Phase transformation behavior and stability of LiNiO2 cathode material for Li-ion batteries obtained from in situ gas analysis and operando X-ray diffraction," *ChemSusChem*, vol. 12, no. 10, pp. 2240–2250, 2019.

[42]   M. Bianchini *et al.*, "From LiNiO2 to Li2NiO3: synthesis, structures and electrochemical mechanisms in Li-rich nickel oxides," *Chemistry of materials*, vol. 32, no. 21, pp. 9211–9227, 2020.

[43]   O. J. Borkiewicz, K. M. Wiaderek, P. J. Chupas, and K. W. Chapman, "Best practices for operando battery experiments: Influences of X-ray experiment design on observed electrochemical reactivity," Jun. 04, 2015, *American Chemical Society*. doi: 10.1021/acs.jpclett.5b00891.

# Supporting Information

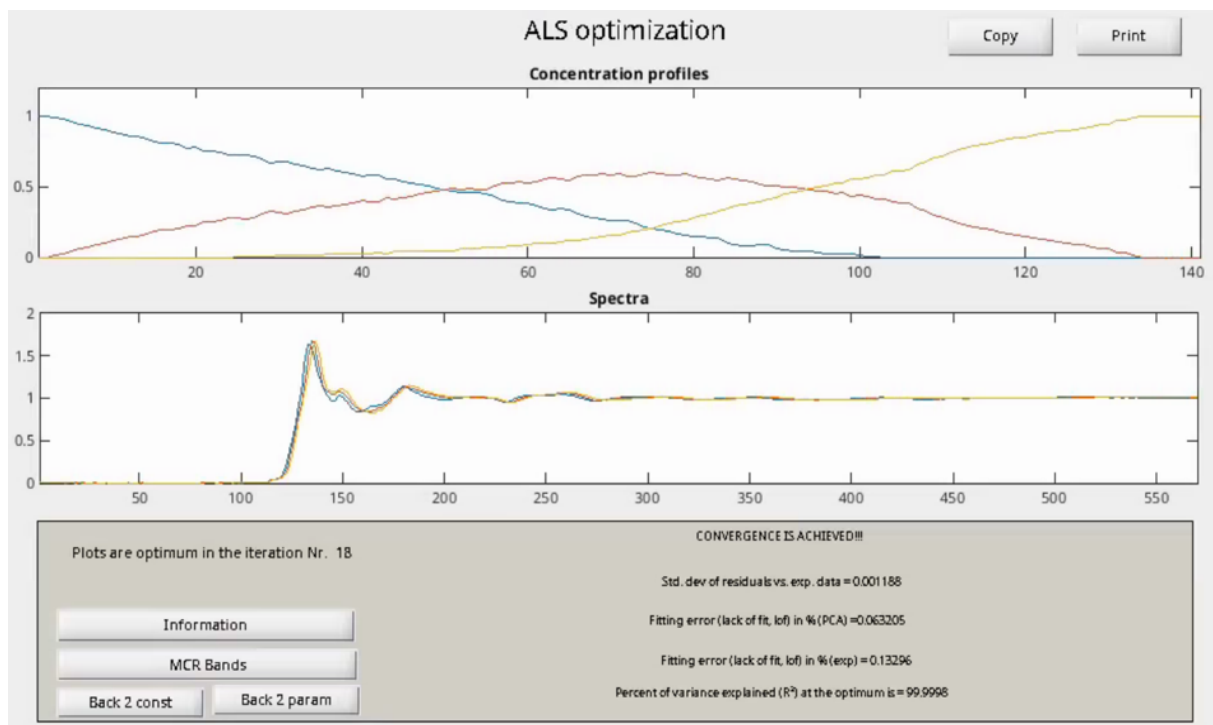

Figure S1: Results of the MCR-ALS analysis on the 141 QX-HC using the dose mask of 0.7-1.0 kGy/s during the charge in farther focused configuration.

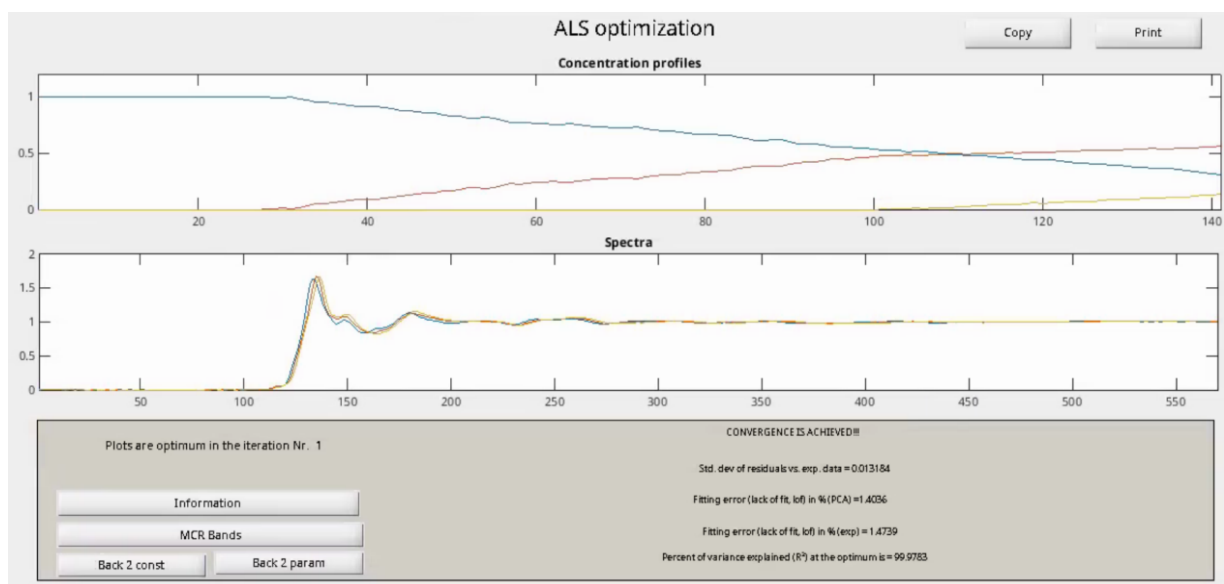

Figure S2: MCR-ALS analysis on the 141 QX-HC using the dose mask of 6.7 – 7 kGy/s during the charge in farther focused configuration. In addition to the previous constraints on the concentration and spectra profile, we have imposed the previous reconstructed Cps as input for the spectra profile.

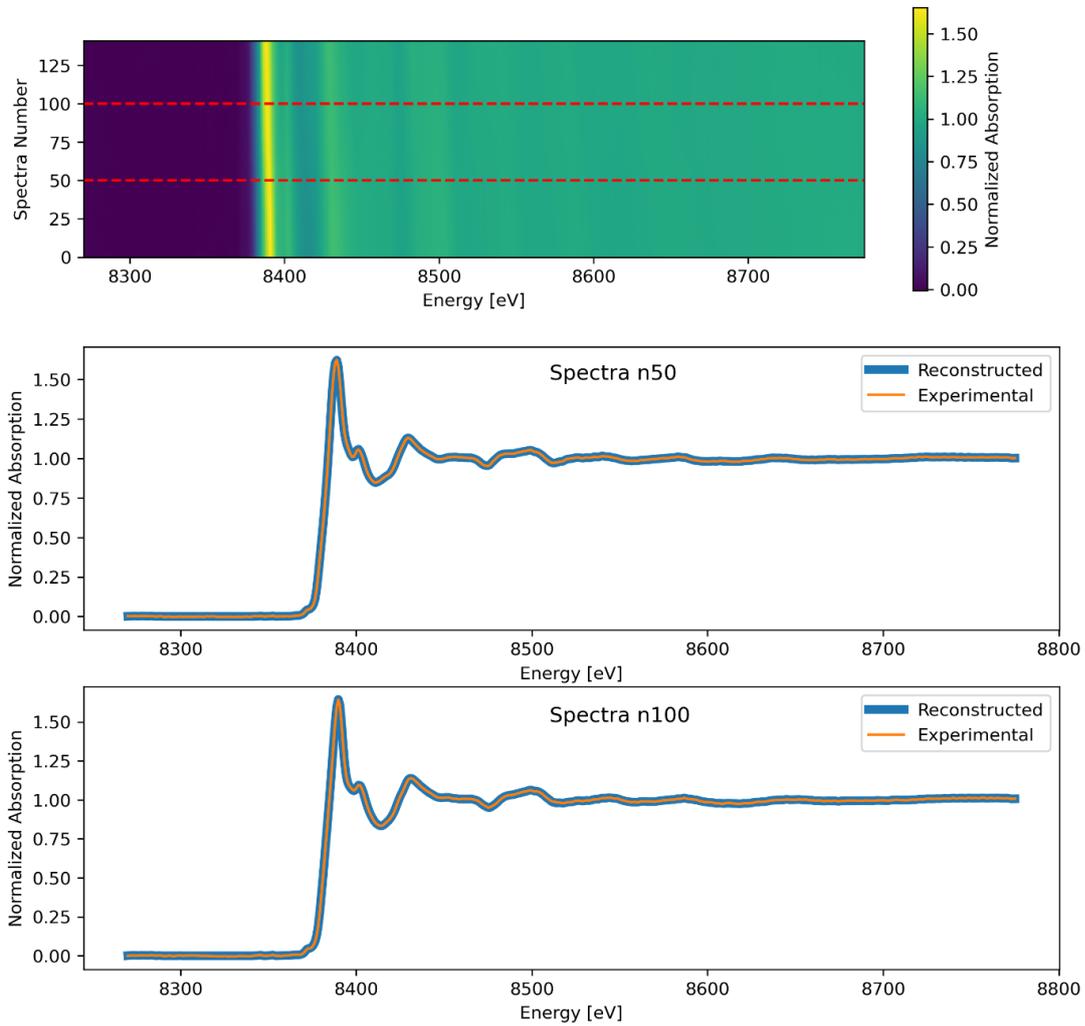

Figure S3: On top: colour map of the averaged spectra obtained during the charge and collected between 0.7 and 1.0 kGy/s for farther focus configuration. The edge shift of the Ni K-edge is clearly visible. Central and bottom panels show a comparison of selected experimental spectra (#50 and #100) with the corresponding linear combination of the reconstructed Cps spectra.

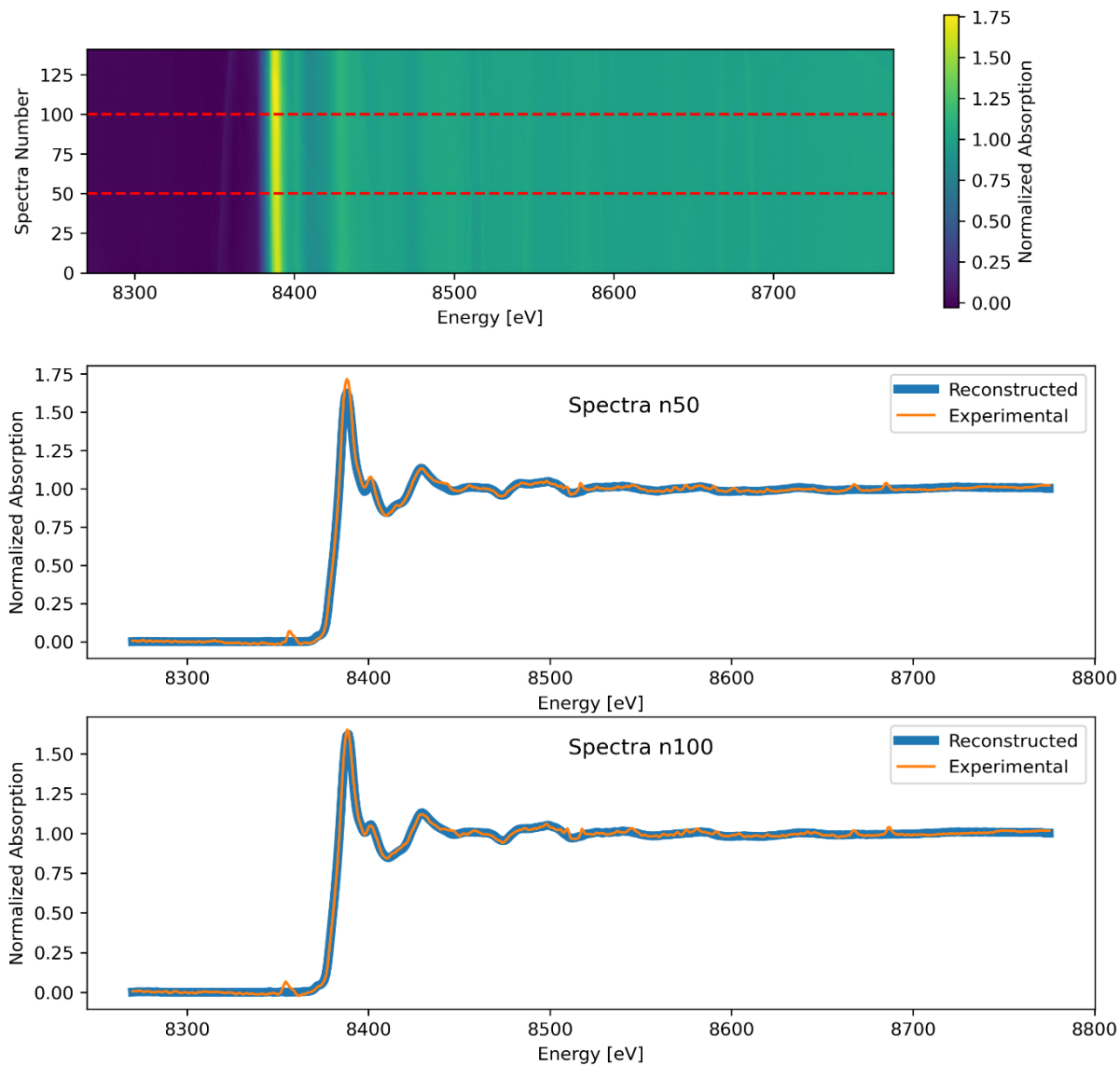

Figure S4: On top: colour map of the averaged spectra obtained during the charge and collected between 6.7 and 7.0 kGy/s for farther focus configuration. Central and bottom panels show a comparison of selected experimental spectra (#50 and #100) with the corresponding linear combination of the reconstructed Cps spectra.

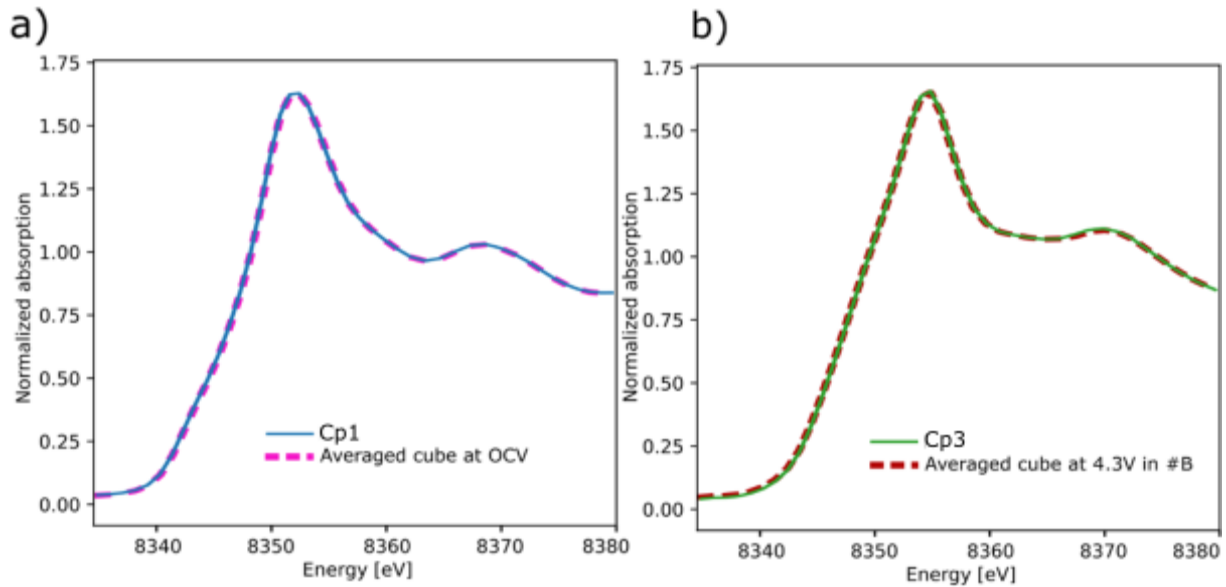

Figure S5: The reconstructed spectra of pure species obtained by MCR-ALS analysis during the charge and collected between 0.7-1.0 kGy/s for the farther focus configuration. The Cp spectra are compared to the averaged spectra of the cube at OCV and at 4.3V in #B.

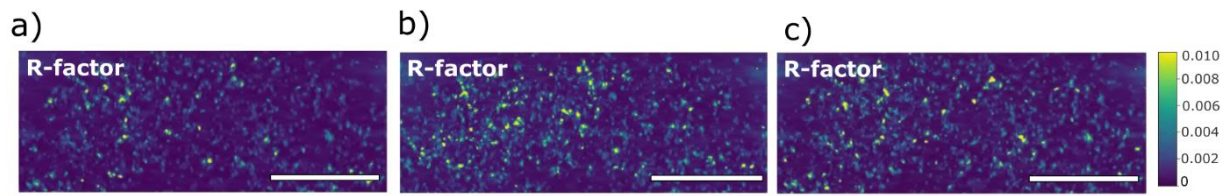

Figure S6: The R-factor maps resulting from the MCR-ALS fit of OCV (a), at 4.3 V in #B (b), and at 4.3 V in #C (c) in the case of farther focus beam configuration. The pixel size is 6.5x6.5µm²; the scale bar is 1 mm. The intense spots observed in the R-factor images were caused by the diffraction peaks of the diamond windows. A certain number of crystal grains of the diamond window were aligned with the X-rays beam, and therefore, it was locally deviated. The camera detected a missing intensity due to the primary extinction, resulting in a peak in the absorption spectra. This localized peak on the final spectra increased the fitting error. However, since the spots are stochastic in nature, integrating over a relatively large number of pixels causes these deviations to average out, and the peaks effectively disappear.

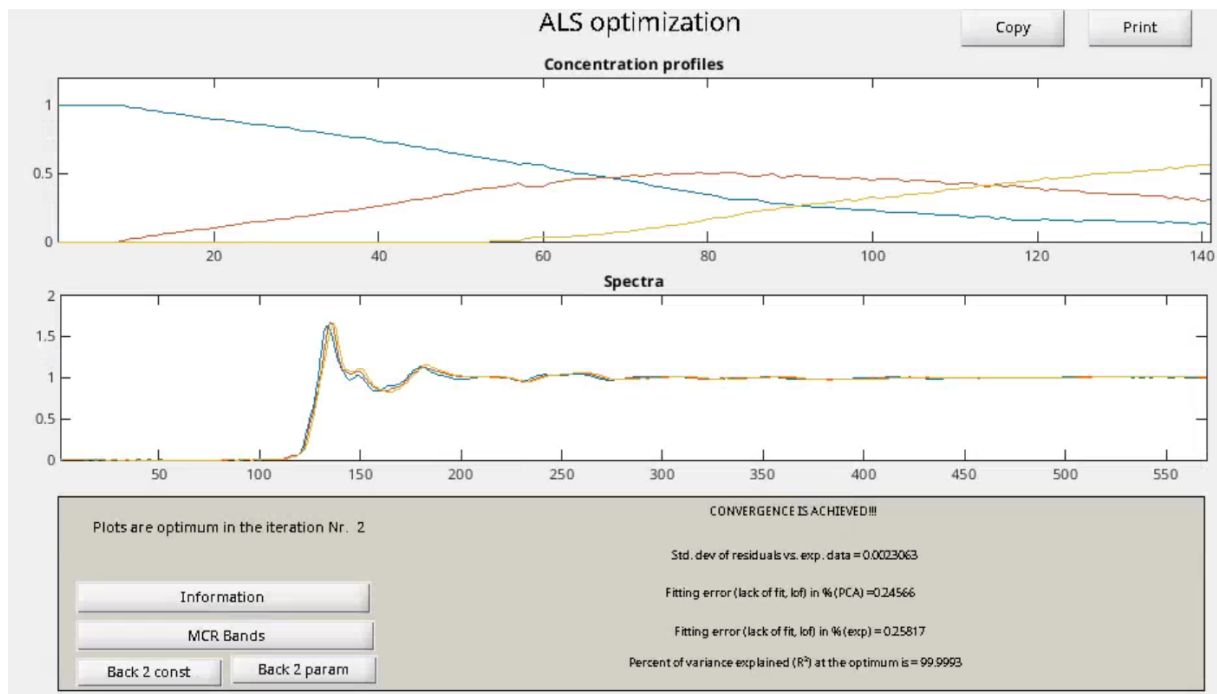

Figure S7: Results of the MCR-ALS analysis on the 141 QX-HC using the dose mask of 0.7-1.0 kGy/s during the charge in closer focus configuration. As input basis for the spectral deconvolution, the XAS spectra of Cp1, Cp2 and Cp3 collected in farther focus configuration were used (Figure 6a).

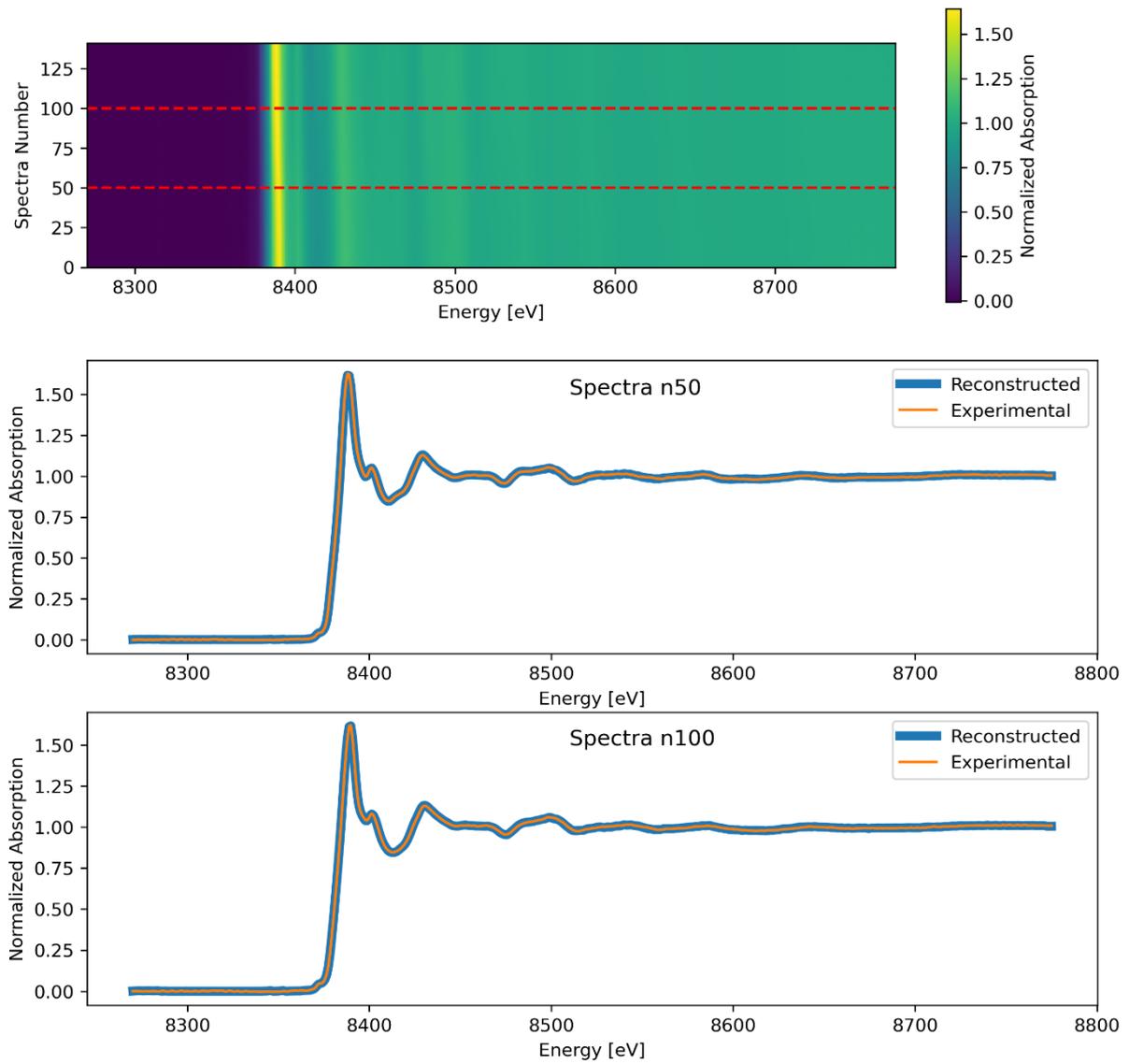

Figure S8: On top: colour map of the averaged spectra obtained during the charge and collected between 0.7 and 1.0 kGy/s for closer focus configuration. Central and bottom panels show a comparison of selected experimental spectra (#50 and #100) with the corresponding linear combination of the reconstructed Cps spectra.

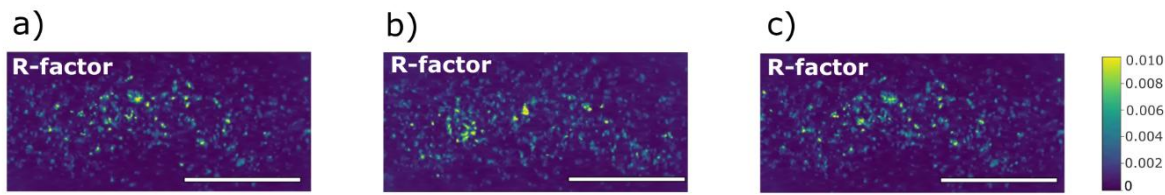

Figure S9: The R-factor maps resulting from the MCR-ALS fit of OCV (a), at 4.3 V in #B (b), and at 4.3 V in #C (c) in the case of closer focus beam configuration. The pixel size is 6.5x6.5µm²; the scale bar is 1 mm. The intense spots observed in the R-factor images were caused by the diffraction peaks of the diamond windows. The nature of the peaks is discussed in Figure S9.

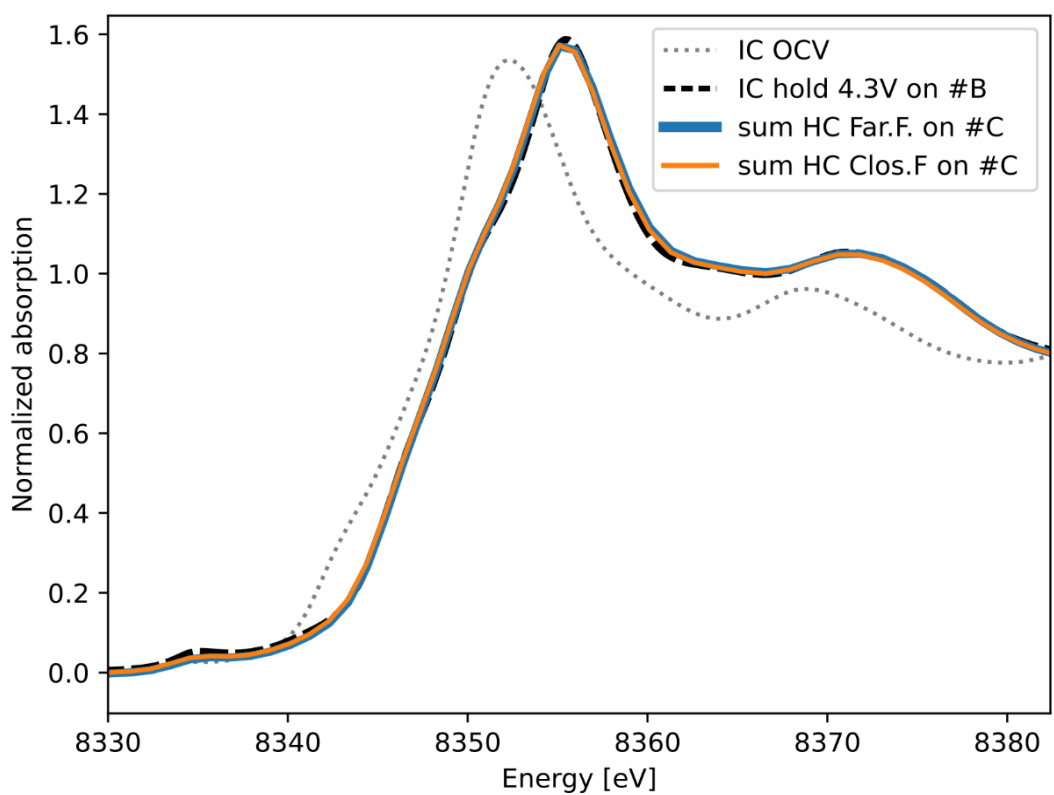

Figure S10: Average of all the pixels spectra of the HC cubes after holding at 4.3 V for 'farther focus configuration' (blue line) and 'closer focus configuration' (orange line). IC spectra acquired at OCV and at hold 4.3V in point #B are displayed as reference.

**Calculation of the dose rate on the LNO electrode, considering the upstream components**

```
from dose_estimation import Beam, Material, Layer, Stacking
```

```python
#------------------------------------------------------------------------------
Diam_window = Layer(name = "Diamond window",
                    solidMaterial = Material(name = "Diamond", formula = "C", density = 3.5),
                    liquidMaterial = None,
                    thickness = 0.145,
                    porosity = 0.0)

#------------------------------------------------------------------------------
# COUNTERELECTRODE
Li_ce = Layer(name = "Counter-electrode",
              solidMaterial = Material(name = "Li", formula = "Li", density = 0.534),
              liquidMaterial = None,
              thickness = 1, # https://www.mtixtl.com/EQ-Lib-LiC60-300.aspx
              porosity = 0)

#------------------------------------------------------------------------------
# CATHODE
cathode = Layer(name = "Cathode",
                solidMaterial = Material(name = "LNO", formula = "LiNiO2", density = 4.716),
                liquidMaterial = Material(name = "electrolyte", formula = "C4H8O3", density = 1.01),
                thickness = 0.01,
                porosity = 0.36)

#------------------------------------------------------------------------------
# SEPARATOR cellgard
separator_c = Layer(name = "separator Celgard",
                    solidMaterial = Material(name = "polypropylene", formula = "C3H6", density = 0.905),  # ? it is made of polypropylene
                    liquidMaterial = Material(name = "electrolyte", formula = "C4H8O3", density = 1.01),
                    thickness = 0.025,
                    porosity = 0.55) # file:///C:/Users/la-porta/Downloads/Evaluation_of_methods_for_the_determination_of_tor.pdf

fx = 1.041e12 # total beam flux measured

xr = Beam(beamEnergy = energ_im,
          photonFlux = fx , # 500 mA
          beamWidth = 1.625 * 2040   ,
          beamHeight = 1.625 * 740   ,
          exposure = 1 ) # we're interested on the dose rate

# BUILD THE CELL (in downstream order, i.e. first the component where the beam enters)
pouch_cell = Stacking([ Diam_window, Li_ce, separator_c,cathode])
#------------------------------------------------------------------------------
# COMPUTE
pouch_cell.get_dose(xr)
#------------------------------------------------------------------------------
# VISUALIZE
fig,ax = pouch_cell.get_visual_dose()
```